\documentclass{aastex631} 

\newcommand{\VLSR}{V_{\rm LSR}}
\newcommand{\FeII}{[\ion{Fe}{2}]}
\newcommand{\SII}{[\ion{S}{2}]}
\newcommand{\OI}{[\ion{O}{1}]}

\newcommand{\kms}{km~s$^{-1}$}
\newcommand{\dotsec}{\rlap.^{''}}
\newcommand{\dotdeg}{\rlap.^{\circ}}

\newcommand{\degree}{^{\circ}}

\shorttitle{Jet and Mass Accretion Variability in DG~Tau}
\shortauthors{Pyo, et al.}

\begin{document}
\title{Ejection Patterns in the DG Tau Jet Over the Last 40 Years: \\Insights into Mass Accretion Variability\footnote{Based on observations obtained at the international Gemini Observatory, a program of NSF's NOIRLab, which is managed by the Association of Universities for Research in Astronomy (AURA) under a cooperative agreement with the National Science Foundation on behalf of the Gemini Observatory partnership.}
}

\author[0000-0002-3273-0804]{Tae-Soo Pyo}
\affiliation{Subaru Telescope, National Astronomical Observatory of Japan, National
Institutes of Natural Sciences, 650 North A`oh\=ok\=u Place, Hilo, HI 96720, USA} 
\affiliation{School of Mathematical and Physical Science, SOKENDAI (The Graduate University for Advanced Studies), Hayama, Kanagawa 240-0193, Japan}
\email{pyo@naoj.org}
\author[0000-0002-4790-7940]{Masahiko Hayashi}
\affiliation{National Astronomical Observatory of Japan, 2-21-1 Osawa, Mitaka, Tokyo 181-8588, Japan}
\affiliation{JSPS Bonn Office, Ahrstr. 58, 53175 Bonn, Germany}
\author[0000-0001-9248-7546]{Michihiro Takami}
\affiliation{Institute of Astronomy and Astrophysics, Academia Sinica, 11F of Astronomy-Mathematics Building, No.1, Sec. 4, Roosevelt Rd, Taipei 10617, Taiwan, R.O.C.}
\author[0000-0002-6881-0574]{Tracy L. Beck}
\affiliation{The Space Telescope Science Institute, 3700 San Martin Dr., Baltimore, MD 21218, USA}

\begin{abstract}
We aim to clarify the link between mass accretion and ejection by analyzing DG Tau's jet observations from optical and near-infrared data spanning 1984 to 2019, alongside photometric variations between 1983 and 2015. We classified 12 moving knot groups among 17 total knot groups based on their constant proper motions and comparable radial velocities. A strong correlation emerges between deprojected flow velocities of the knots and the photometric magnitudes of DG Tau. From 1983 to 1995, as the deprojected ejection velocities surged from $\sim$~273~$\pm$~15 \kms\ to $\sim$~427~$\pm$~16 \kms\, the photometric magnitudes ($V$) concurrently brightened from 12.3 to 11.4. Notably, when DG Tau became brighter than 12.2 in the $V$ band, its (\bv) color shifted bluer than its intrinsic color range of K5 to M0. 
During this period, the launching point of the jet in the protoplanetary disk moved closer to 0.06 AU from the star in 1995.  
Following a $V$ magnitude drop from 11.7 to 13.4 in 1998, the star may have experienced significant extinction due to a dust wall created by the disk wind during the ejection of the high-velocity knot in 1999.
Since then, the magnitude became fainter than 12.2, the (\bv) and (\vr) colors became redder, and the deprojected velocities consistently remained below 200 \kms. The launching point of the jet then moved away to $\sim$ 0.45 AU by 2008.
The prevailing factor influencing photometric magnitude appears to be the active mass accretion causing the variable mass ejection velocities.
\end{abstract}

\section{INTRODUCTION}\label{sec:intro}

Brightness variability is intrinsic to young stellar objects (YSOs). 
Various phenomena, such as outbursts, mass ejection (loss), and mass accretion, drive this variability.
For example, FUors display a 4 to 6 mag increase in visual magnitude, often attributed to outbursts resulting from significant upticks in the mass accretion rate \citep{HK1996}. 
Numerous studies have investigated the correlation between mass ejection events associated with knots in YSO jets and changes in photometric magnitude.
\citet{Garufi2019} noted that two knots in the jet of RY~Tau were ejected during local photospheric peaks in the $V$ magnitude. 
On the other hand, \citet{Babina2016} discovered a correlation between stellar magnitude and the radial velocity of the wind from RY~Tau.
\citet{Takami2020} identified a potential correlation between the ejections of four knots and the dimming events observed in the optical photometric data of RW~Aur. 

DG Tau is one of the first young stellar objects for which optical jets were discovered in the early 1980s \citep{Mundt1983}. 
\citet{Mundt1983} reported that the DG~Tau jet (also known as HH 158)  extends up to $\sim$ 10$\arcsec$ in the southwest direction from the star (PA $\sim$ 225$^{\degree}$). 
\citet{EM98} estimated the proper motions and radial velocities for the  four knots.
Since their work, only two papers have observed the entire length of the jet \citep{Whelan2004, Oh2015}. 
\citet{Oh2015} demonstrated that the jet extended up to 20$\arcsec$ by 2014. Other studies have focused their observations primarily around the star within a span of several arcseconds \citep[e.g.][]{Bacciotti2000,Lavalley2000,Agra2011,White2014_1,Maurri2014,Takami2023}.

\citet{Pyo03} reported from observations of the \FeII\ $\lambda$1.644 $\mu$m line that the DG~Tau jet possesses two distinct velocity components characterized by their radial velocities ($V_r$) and spatial distributions. 
The high-velocity component (HVC) has velocities of $V_{\rm LSR}~=~-$220~$\pm$~50 \kms\  and an intensity peak at $\sim$0$\dotsec$75 away from the star\footnote{The stellar radial velocity is $V_{\rm LSR}~=~+6.4$ km s$^{-1}$.}, while the low-velocity component (LVC) is measured at $-$100~$\pm$~100 \kms\ and shows an intensity peak at $\sim$0$\dotsec$45 from the star. 
They concluded that the HVC is associated with a fast, well-collimated jet originating from either the innermost disk, the star-disk interface, or the star surface. In contrast, the LVC represents a broader wind, likely arising from the outer radii near the inner edge of the disk. 
\citet{White2014_1} and \citet{Agra2011} also provided the position velocity diagrams (PVDs) of the \FeII\ $\lambda$ 1.644 $\mu$m line, acquired on 11th November and 15th October 2005, respectively.
\citet{White2014_1} identified two velocity components: one located at $\sim$0$\dotsec$4 with a velocity of $-$180 \kms\ and another at $\sim$1$\dotsec$24 with a velocity of $-$250 \kms. 
These components correspond to their knots B and C, which we will refer to as knots WB and WC, respectively, in this paper.

Notable peaks of this high-velocity flow, surpassing $-$200 \kms, were observed from 1988 to 2002, as evidenced in Figure 1\textit{o} of \citet{Hamann1994} (observed in February 1990), Figures 5 and 7 of \citet{HEG95} (observed on January 9, 1988), and Figures 2\textit{a} and 2\textit{b} of \citet{Whelan2004} (observed on September 17, 2002). 
The PVDs by \citet{Maurri2014} also revealed dominant emissions in the $|V_r| >$ 200 \kms\ range from $\sim$~0$\dotsec$75 (knot A1) to 5$\arcsec$ (knot B1) in January 1990. 
This high-velocity component was similarly noted in the \FeII\ emissions of \citet{Takami2002}, \citet{Pyo03}, and \citet{White2014_1} for the period between 2000 and 2005. 
However, the flux levels at $|V_r| >$ 200 \kms\ were minimal in Figure 1 of \citet{Chou2013}, which detailed line profiles of \SII\ $\lambda$6731 \AA, \OI\ $\lambda$6300 \AA, and \FeII\ $\lambda$7155 \AA\ obtained from October to November 2010. 
Furthermore, \citet{Takami2023} reported the presence of a single collimated jet with $|V_r|=$ 120$-$160 \kms\ over 2013-2019. This jet exhibited a notably reduced velocity when compared to previous observations.
This indicates that the high-velocity flow exceeding $|V_r| >$ 200 \kms\ was present from 1988 to 2005 but has been weakened since 2006. 
This variation implies changes in the inner disk structure, given that the high-velocity flow is likely associated with the inner boundary layer between the stellar surface and the inner edge of the circumstellar disk \citep{Pyo03}.

Knots may form due to internal shocks between ejecta of varying velocities, exhibiting periodic spacing. 
The ejection of knots in HH 158 has been extensively studied. 
\citet{Pyo03} proposed a knot ejection cycle of 5 years, determined from the proper motions of the knots. 
This was later corroborated by \citet{Rod2012} using radio, optical, and X-ray data.  \citet{Agra2011} proposed a shorter period of 2.5 years, deduced from the presence of an intermediary knot and assuming a typical proper motion ranging from 0$\dotsec$27 to 0$\dotsec$34 yr$^{-1}$. 
\citet{White2014_1} suggested that the ejection period might be variable, citing the lack of a new knot for 4 years between November 2005 and November 2009, as well as the atypical proper motion of 0$\dotsec$17 yr$^{-1}$ observed for knot WB.

Beyond the variability associated with knots, T~Tauri stars show significant photometric variability.
\citet{Grankin2007} summarized the optical photometric variations of DG Tau over a period of 20 years, spanning from 1984 to 2004.
They reported that DG~Tau experienced large changes in its mean brightness ($\sigma_{V_{\rm m}}$) over this timescale.
Specifically, DG Tau's brightness gradually increased, fluctuating from 12.5 to 11.5 mag ($\Delta V_{\rm m} = -$1.0) between 1984 and 1996. 
There was a pronounced drop in the mean brightness from 11.71 to 13.31 mag ($\Delta V_{\rm m} = +$1.6) between 1997 and 1998.

In this paper, we provide a comprehensive summary of knot ejection events in DG~Tau over the past 40 years, detailing their proper motions and radial velocities, and discussing their correlation with the star's photometric variability.

\section{OBSERVATIONS AND DATA REDUCTION}\label{sec:add_feII}

We incorporated new data of the \FeII\ $\lambda$1.257 $\mu$m emission line observed on 13th February 2007 (UT), using the Near-Infrared Integral Field Spectrometer (NIFS) mounted on the Gemini-North Telescope atop Mauna Kea, Hawai`i. 
NIFS, an image-slicing integral field unit, is complemented by the adaptive optics (AO) system ALTAIR\footnote{ALTtitude conjugate Adaptive optics for the InfraRed}.
It offers a high spatial resolution utilizing either a natural guide star (NGS) or a laser guide star (LGS) \citep{McGregor2002}.

Given its sufficient brightness with $R$ $=$ 12.28 mag, we employed DG Tau as the NGS for AO correction.
After the AO correction, the full width at half maximum (FWHM) of the stellar image was $\sim0\farcs14$. 
The Integral Field Unit (IFU) offered a spectral resolution of $R$~$\sim$~5000, covering a 3$''\times$3$''$ field of view at a pixel scale of 0$\farcs$1$\times$0$\farcs$04.
The observations were conducted under photometric conditions, with the average natural seeing ranging between 0$\farcs$6 and 0$\farcs$85. 
Further details on the observations and data reduction can be found in \citet{Pyo14} and \citet{Beck2008}. 
The reduced data was subsequently re-sampled to a square grid with a pixel scale of 0$\farcs$04 $\times$ 0$\farcs$04 pixels.

\section{RESULTS} \label{setion:results}
\subsection{Knots and velocity structure in the new data}

\begin{figure*}
\epsscale{1.0}
\plotone{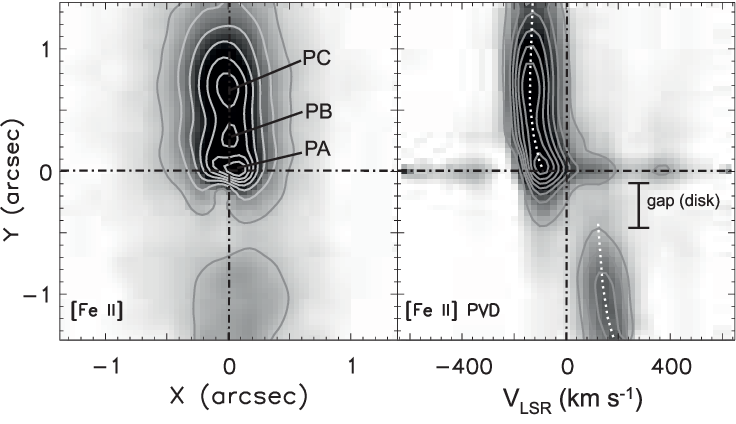}
\caption{(\textit{left})The image on the left shows the \FeII\ $\lambda$1.257 $\mu$m emission after removing the continuum. The three knots along the jet, PA, PB, and PC, are marked and labeled. The field of view is $2\farcs6 \times 2\farcs6$, and the central star, DG Tau, is located at (X,Y)=(0,0). The vertical and horizontal dash-dotted lines indicate the star position. The emission line was integrated over $\pm$ 600 \kms. The contour levels are 10$\sigma$, 20$\sigma$, 30$\sigma$, 41$\sigma$, 53$\sigma$, 66$\sigma$, and 81$\sigma$ above the sky level where $\sigma$ is the standard deviation of the sky area.  The upward direction (Y $\ge$ 0$\arcsec$) indicates the direction of the blue-shifted jet with a position angle (PA) of 222$\degree$. The red-shifted jet is moving in the opposite direction (Y $\le$ 0$\arcsec$) from the blue-shifted jet, leaving a gap of about 1 arc second at the star position. The gap is the part obscured by the disk \citep{Pyo03}.
(\textit{right})The position-velocity diagram on the right shows the same emission.  The systemic radial velocity is $\VLSR = +$6.4 \kms\ \citep{Kimura1996}. The vertical dash-dotted line marks V$_{LSR} =$ 0 \kms. The peak velocity along the jet is indicated with the dotted line. The contour levels are 10$\sigma$, 15$\sigma$, 21$\sigma$, 27$\sigma$, 35$\sigma$, 43$\sigma$ and 52$\sigma$ above the background level where the $\sigma$ is the standard deviation of the background. It was observed using ALTAIR with NIFS/Gemini on February 13, 2007 (UT). The spatial resolution was $0\farcs14$, and the spectral resolution was 60 \kms\ ($R~\sim$ 5000). 
\label{fig:FeII}
}
\end{figure*}

The left panel of Figure~\ref{fig:FeII} shows the continuum-subtracted image of the \FeII\ emission, integrated over the velocity range of $-$600 \kms\ to +600 \kms. 
The positive direction of the vertical axis aligns with the blue-shifted jet direction, at PA $=$ 222$\degree$. 
It reveals a well-collimated jet. 
The red-shifted jet becomes visible at Y $<-$0$\dotsec$6, traversing a gap likely induced by the protoplanetary disk \citep{Pyo03,Agra2011,White2014_1}.
Notably, the intensity of the red-shifted \FeII\ emission is 4--5 times less than the peak brightness of the blue-shifted jet located at Y = 0\farcs72.

The blue-shifted jet reveals three intensity peaks at positions 0$\arcsec$, 0$\dotsec$30, and 0$\dotsec$72, which we have designated as knots PA, PB, and PC, respectively. 
We compared these to the knots A, B, and C of \citet{White2014_1} (hereinafter WA, WB, and WC). \citet{White2014_1} observed a comparable area as we did three times between 2005 and 2009, while we conducted our observations in 2007.
The position of the knot PA is situated closely to the star's position within a spatial resolution of 0$\farcs$14. Consequently, it may be influenced by residual effects following the subtraction of the star's continuum. Previous work by \citet{White2014_1} identified a stationary knot WA, situated approximately at 0$\farcs$2, spanning the years 2006 to 2010.
Considering the aforementioned reasons, we have excluded knots PA and WA from the list of moving knots.
Knot PB, found at 0$\dotsec$30, when hypothesized to align with knot WA, reveals a proper motion of 0.06 $\arcsec$ yr$^{-1}$, significantly smaller than other identified knots. 
Figure 10 in \citet{White2014_1} does not exhibit any knot structure roughly 0$\dotsec$47 away from the star either.
This obscures the relationship between knots WA and PB. 
Finally, the location of knot PC overlaps with the trajectory of knot WB from \citet{White2014_1}. 
It indicates that the knot PC belongs to the knot WB.

The right panel of Figure~\ref{fig:FeII} presents the PVD of the jet, detailing the velocity characteristics of the \FeII\ $\lambda$1.257 $\mu$m emission. 
The blue-shifted \FeII\ emission spans velocities from $-$215 to 0 \kms, delineating a clear boundary at $\sim-$220 \kms, 
aligning with the LVC of the \FeII\ $\lambda$1.644 $\mu$m emission described by \citet{Pyo03}. 
The peak velocity fluctuates between $-$150 to $-$100 \kms, with no discernible higher velocity component exceeding $-$200 \kms. 
The absence of HVC contrasts sharply with Figure~1 from \citet{Pyo03}, which displays a pronounced HVC extending to $\sim-$300 \kms\, peaking at $\VLSR = -$220 \kms. 
\citet{Agra2011} and \citet{White2014_1} presented PVDs of the \FeII\ $\lambda$ 1.644 $\mu$m line, obtained in 2005, roughly two years prior to our \FeII\ observations. 
When considering the proper motion of the knots and the interval between observations, the features they observed within 1$\arcsec$ of the star largely align with ours. 
The PVD in Figure 4 of \citet{Agra2011} displays a broader velocity range, extending to $-$400 \kms\, attributed to a lower wavelength resolution of $R\sim$ 3,000. 
Meanwhile, Figure 15 of \citet{White2014_1} reveals the emission spanning up to $-$280 \kms\ in their NIFS dataset. 
Their data shows the emission stretching to velocities beyond $-$200 \kms.
In contrast, the PVDs taken after 2013 by \citet{Takami2023} echo the velocity features of ours, indicating a single lower-velocity flow at $V_{\rm LSR} =$ $-170$ to $-140$ \kms\ within 2$\arcsec$ from the star.

\subsection{Identification of knots}\label{subsec:ID_knots}

\begin{figure*}
\epsscale{1.1}
\plotone{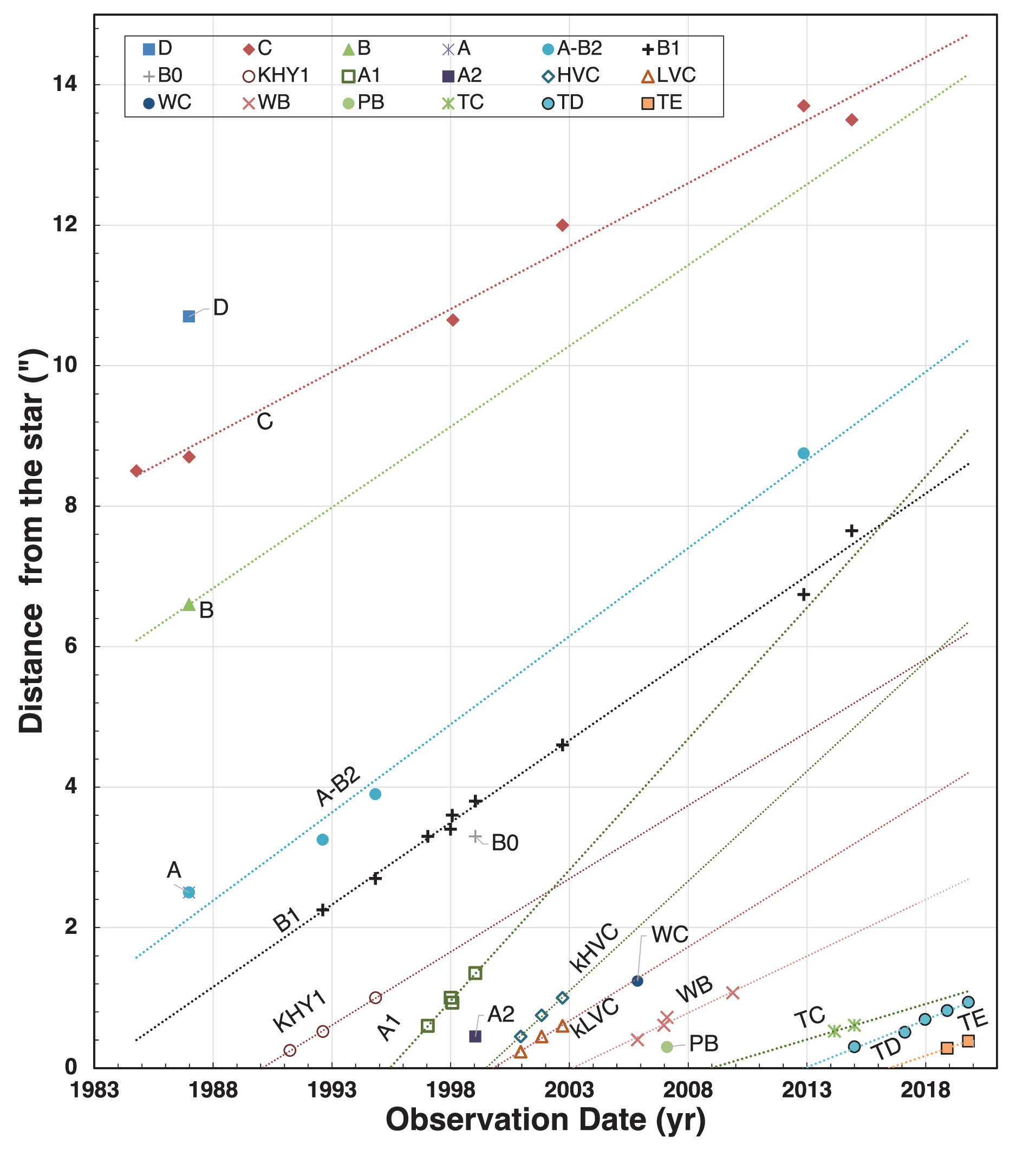}
\caption{Observed knots along the DG~Tau jet from 1984 to 2019. The dotted lines indicate the proper motions of the 12 knot groups. 
\label{fig:knots}
}
\end{figure*}

Table~\ref{table:knots} and Figure~\ref{fig:knots} compile all the moving knots observed in the DG Tau jet from 1984 to 2019, as reported in the literature. 
A total of 51 knots have been documented. 
These knots have been observed in optical and near-infrared forbidden emission lines. 
Notably, over half of the knots are within 2$\arcsec$ from the star (approximately 51\% within 1$\arcsec$ and about 57\% within 2$\arcsec$).
This reminds us that most observations have been concentrated on detecting knots close to the star with a narrow field of view using high spatial resolutions with AO.

Assuming constant velocity trajectories following ejection from the star, we have categorized these observations into 17 unique knot groups, each defined by a consistent proper motion (see Table~\ref{table:knots}).
The knots associated with the HVC and LVC in the study by \citet{Pyo03} are denoted as kHVC and kLVC, respectively, to prevent any ambiguity with general high- and low-velocity components.
For several knots, namely B, B0, A2, WC, and PB, subsequent observations did not identify their presence further down the jet, suggesting they might have been dispersed, merged with other knots, or no longer be shocked as they traveled downstream. 
Our analysis will concentrate on knot groups that exhibit long-term stability. 
Of the 17 knot groups, we estimated the proper motions for 12 of them as discussed below.
The 12 moving knot groups are listed in Table~\ref{tab:pm}.

\citet{Mundt1987} reported for the first time an optical jet from DG Tau, which corresponds to knot C of \citet{EM98} who identified four knots, labeled A, B, C, and D, in their \SII\ images. 
While knots B and C were distinctly identified by peaks enclosed by closed contours, knots A and D lacked such defining characteristics. 
Knot D is positioned ahead of the luminous knot C, making it seem like an extension of knot C without an individual peak. 
Extrapolating from its observed proper motion of $\sim$~0\farcs32~yr$^{-1}$, as reported by \citet{EM98}, the projected location of knot D in 2015 would be $\sim$~20$\arcsec$ distant from the star. 
Yet, \citet{Oh2015} observed a continuous emission trail from the central star to knot C, not discerning a distinct knot corresponding to D. 
It is plausible to deduce that knot D characterizes the jet's leading edge and invariably precedes knot C.
Of the four knots, knot C was the most distinct at the jet's forefront from 1984 to 2014, demonstrating a proper motion of 0\farcs18~$\pm$~0\farcs01~yr$^{-1}$.
Knot B, a faint peak observed by \citet{EM98} from 1983 to 1990, was not distinctly identified as an independent knot in subsequent long-slit observations after 2002. 
Using the proper motion measurement of 0\farcs23~$\pm$~0\farcs02 yr$^{-1}$ provided by \citet{EM98}, it appears that knot B is converging with Knot C, which has a proper motion of approximately 0\farcs18 yr$^{-1}$. 
In the 2002 observation by \citet{Whelan2004}, knot C spanned over 5$\arcsec$ (extending between approximately 10$\arcsec$ and 15$\arcsec$). 
This extension is likely due to the merging of knots B and C, as the predicted position for knot B was around $\sim$~10\farcs2. 
Since then, the separation between knots B and C has remained under 1\farcs2. 
As a result, knot B has become indistinct, especially in proximity to the brighter knot C, particularly in the low spatial resolution observations (with a seeing of roughly 2$\arcsec$) by \citet{Oh2015}.
Knot A was located 2\farcs5 from the star in the images of \citet{EM98}.
Meanwhile, \citet{SB93} and \citet{Lavalley1997} identified two knots, which were labeled as B1 and B2, in their \SII\ PVDs.
The position of knot A well matches the trail of knot B2, as shown in Figure~\ref{fig:knots}.
We thus posit that knots A and B2 are indistinguishable and henceforth refer to them as knot A-B2.
The proper motion of knot A-B2 is 0\farcs25 yr$^{-1}$.

Knot B1 has been observed multiple times by many observers from 1992.62 to 2014.89, exhibiting a proper motion of 0\farcs23~$\pm$~0\farcs01 yr$^{-1}$.
Knot B0 was located at 3\farcs3, which was 0\farcs5 behind knot B1, in 1999 \citep{Maurri2014}.
It exhibited a greater radial velocity of $-$255 \kms\ compared to the $-$210 \kms\ of knot B1.
In 2002, \citet{Whelan2004} showed a long knot at 4\farcs6, called knot B, whose radial velocity varies from $-$240 \kms\ at 4$\arcsec$ to $-$200 \kms\ at 6$\arcsec$. 
This velocity gradient along the long knot is probably due to the merge effect of the faster knot B0 at the behind of the knot B1. The two close knots could not be separated with low spatial resolution $\ge$ 1$\arcsec$. 
Any fainter knots, should they exist, interspersed between knots B1, B2, and C remain unresolved due to a spatial resolution of $\sim2''$, as depicted in Figure 3 by \citet{Oh2015}.

A knot at 0\farcs25 found by \citet{Kepner1993}, knot C by \citet{SB93}, and the extension of the high-velocity branch by \citet{Lavalley1997} align well in a constant proper motion, as illustrated in Figure~\ref{fig:knots}. 
These are collectively categorized as knot KHY1 in Table~\ref{table:knots} with the proper motion of 0\farcs21~$\pm$~0\farcs01 yr$^{-1}$.

\citet{Bacciotti2000} identified the two distinct features A1 at 1\farcs45 and A2 at 0\farcs75 in 1999.
They interpreted the positions of A1 and A2 as the edges of the extended emission structure, rather than peak positions. 
\citet{Maurri2014} presented position-velocity diagrams from seven contiguous slits, covering an extended area up to 5\farcs5 from the star. 
The peak position for knot A1 was identified slightly closer to the star, at 1\farcs35. 
Knot A2 is associated with diffuse continuous jet emission extended to 0\farcs75 from the star.
We selected the midpoint at $\sim$~0\farcs45 for knot A1, with an uncertainty of $\pm$0\farcs2.
Our analysis indicated that the peak position of knot A1 aligns well with a constant proper motion when considering three subsequent observations: a knot structure peak at 0\farcs6 in January 1997, 1\farcs0 in December 1997 as reported by \citet{Dougados2000}, and the high-velocity peak at 0\farcs93 in January 1998 by \citet{Lavalley2000}. 
The radial velocities measured in these studies were consistently similar, averaging around $-$347~$\pm$~5 \kms. As a result, we categorized these observations under knot A1.

For the knots kHVC and kLVC initially identified by \citet{Pyo03} in the \FeII\ $\lambda$ 1.644 $\mu$m line, we observed a consistent match in locations and radial velocities with the knots identified by \citet{Takami2002} in their Figure~1 and by \citet{Whelan2004} in their Figure~3\textit{b}.
The calculated proper motions for knots kHVC and kLVC are 
0\farcs31 yr$^{-1}$ and 0\farcs21 yr$^{-1}$, respectively, based on the data from these three observations.

Knot WC is situated along the path of kLVC in Figure~\ref{fig:knots}. 
However, its radial velocity of $-$250 \kms\ aligns more closely with that of kHVC. 
If knot WC were a successor to kHVC, its peak would be expected at 1\farcs98 in 2005, placing it outside the field of view of NIFS. 
Instead, knot WC was observed at 1\farcs24, near the edge, and appeared to extend beyond that point. 
One possibility is that WC represents the tail end of kHVC, truncated by the field of view. 
Alternatively, knot WC might be a distinct high-velocity knot with a velocity of $-$250 \kms. 
Given these uncertainties, we have chosen not to incorporate WC into either kHVC or kLVC.

Knot WB, referred to as knot B by \citet{White2014_1}, was documented in three separate instances by \citet{White2014_1}. 
In this paper, we have included knot PC at 0\farcs72 within the knot WB group due to its alignment with knot WB's trajectory, as detailed in Section $\S$\ref{sec:add_feII}. 
Knot PB was identified in this study (see Section $\S$\ref{sec:add_feII}). 
No subsequent observations have captured a corresponding knot since our initial identification, and as a result, we have chosen not to consider it further.

\citet{Takami2023} recently identified knots labeled A, B, C, D, and E. 
This paper refers to them as TA, TB, TC, TD, and TE, respectively. 
Among these, knots TA and TB vanished by 2017 and 2014, respectively. 
Earlier observations did not detect these two knots either.
We hence have chosen not to consider them in the current study. 
The remaining knots, TC, TD, and TE exhibit distinct peaks and discernible proper motions. 
Consequently, we have included them in the current study.

\begin{figure*}
\epsscale{1.1}
\plotone{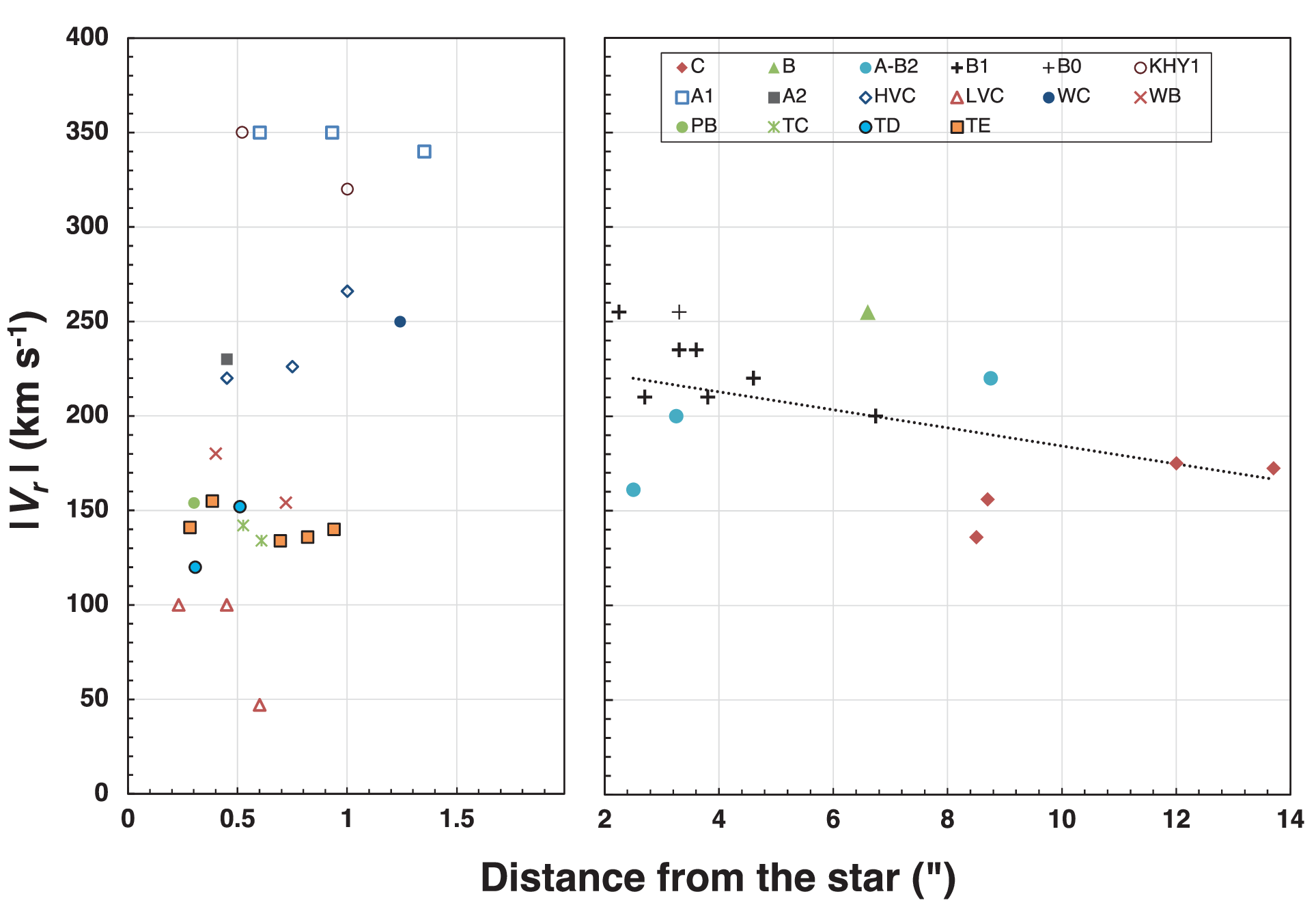}
\caption{Absolute radial velocities ($|V_r|$) of the knots along the DG Tau jet.
The crowded area within 2$\arcsec$ is zoomed (left side) to show the knots in detail. 
The dotted line shows the deceleration trend ( $-$4.8 \kms\ arcsec$^{-1}$) of $|V_r|$ for the knots located further than 2$\arcsec$ away from the star.
\label{fig:Vr}
}
\end{figure*}

Figure~\ref{fig:Vr} shows the radial velocities of the knots plotted against their distances from the star. 
Out of the 51 knots observed, radial velocities were determined for 41 of them. 
All recorded radial velocities are presented as absolute values, $| V_r |$, relative to DG~Tau's radial velocity of $\VLSR = +$6.4 \kms\ \citep{Kimura1996}.
The accuracy of the radial velocity for each knot is within a few ten \kms.

\begin{figure*}
\epsscale{1.0}
\plotone{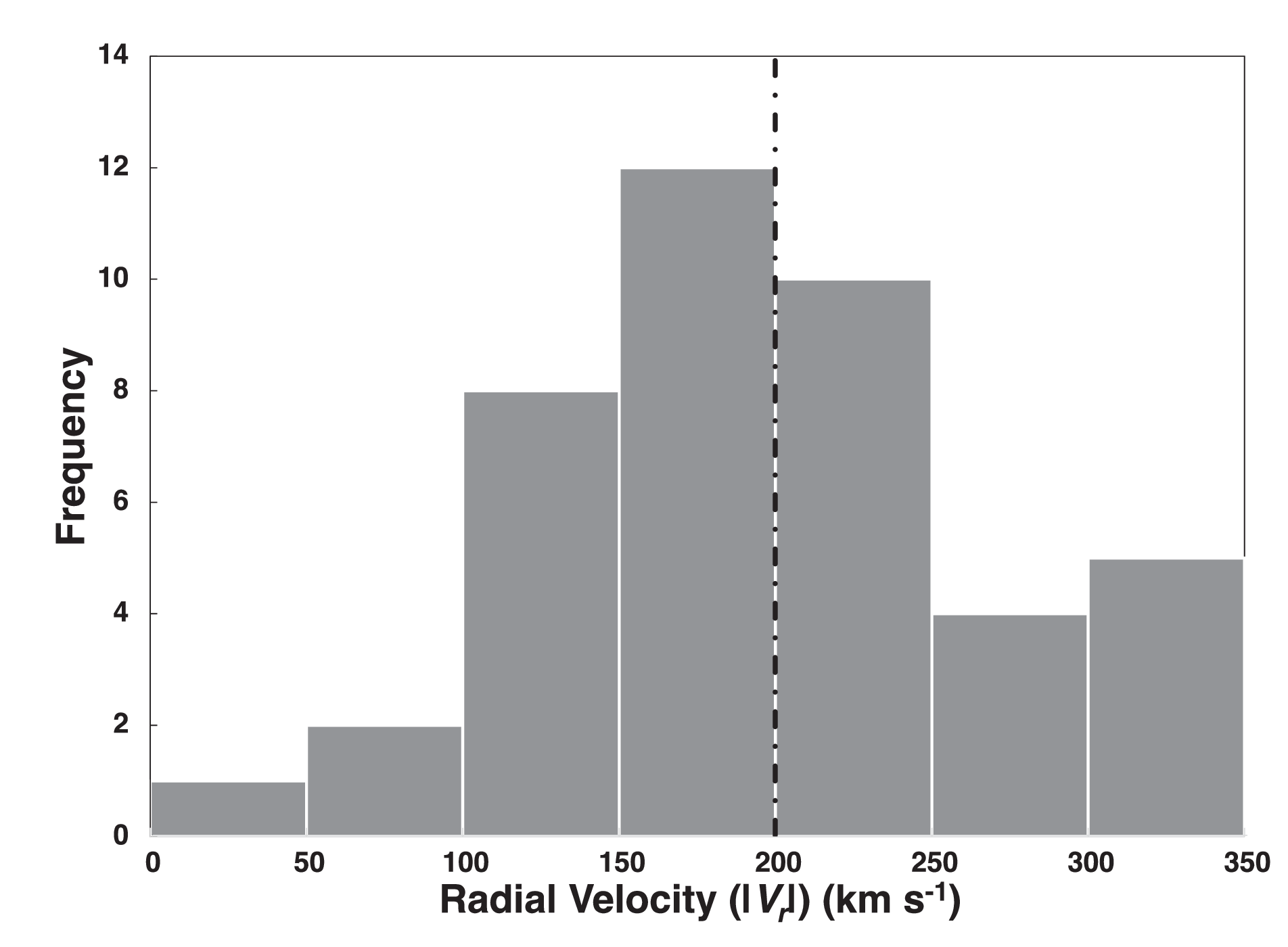}
\caption{Histogram displaying the radial velocities ($|V_r |$) of the knots in the DG~Tau jet. The average velocity is 199~$\pm$~73 \kms\, denoted by a vertical dash-dotted line. Twenty-one knots ($\sim$ 51\%) exhibit velocities within the range of 150 - 250 \kms.
\label{fig:Vr_hist}
}
\end{figure*}

Within 1$''$ of the star, knot velocities span a wide range, from 47 \kms\ up to 350 \kms. 
Conversely, at distances exceeding 2$''$, we observed velocities predominantly clustered around the mid-range value of 206~$\pm$~37 \kms. 
Figure~\ref{fig:Vr_hist} displays the distribution of $|V_r |$ for all 41 observations. 
Notably, 51\% (21 out of 41) of these observations indicate velocities ranging between 150 and 250 \kms, with a mean velocity of 199~$\pm$~73~\kms.
This velocity distribution implies that knots with either very high or very low velocities are confined to within 2$''$ of the star. 
Knots related KHY1 and A1 have very high radial velocities over 300 \kms\ . Their locations are expected around 2$\dotsec$6$-$2$\dotsec$7 from the star on 2002.71. However, \citet{Whelan2004} did not show any knot-like structures around that position and only showed weak diffuse emissions.
Knot kLVC showed the lowest radial velocity, which was no longer discernible in the data of \citet{White2014_1} on 2005.87.
The vanishing of knot TA within $\sim$~2$''$ was reported by \citet{Takami2023}.
Further investigation will be required to provide explanations for these instances of knot disappearance.

The radial velocities of knots beyond 2$''$ decline at a rate of $-$4.8 \kms~arcsec$^{-1}$ while the velocity gradient is much smaller as $-$0.3 \kms~arcsec$^{-1}$ including the inner knots within 2$\arcsec$.
Each knot displays, however, unique characteristics.
Knot B1 exhibits a pronounced deceleration trend of $-$5.4 \kms~arcsec$^{-1}$ spanning from 2\farcs3 to 6\farcs7.
Knot C manifests a subtle increase in speed, transitioning from 136 \kms\ at 8\farcs5 to 172 \kms\ at 13\farcs7.
Likewise, knot A-B2 demonstrates a mild velocity ascent, climbing from 200 \kms\ at 3\farcs3 to 220 \kms\ at 8\farcs8.

\section{EJECTION DATES, PROPER MOTIONS, VELOCITIES, AND INCLINATION ANGLES of TWELVE KNOT GROUPS}\label{sec:pm}

In Table~\ref{tab:pm}, we provide a summary of the observations for the 12 knot groups observed multiple times. 
This table lists the ejection dates, proper motions, average radial velocities ($| \bar{V}_r |$), tangential velocities ($V_t$), deprojected knot velocities ($V_f$), and inclination angles ($I$) relative to the line of sight. 
The proper motion for each knot group was determined using least squares fitting. 
Based on these values, we deduced the ejection dates of the knots from the star position.
Additionally, we computed the average radial velocity for each knot group. 
To transform the proper motion into tangential velocity ($V_t$), we employed the distance to the TMC, which is 141~$\pm$~7 pc, as given by the GAIA DR2 data \citep{Zucker2019}. 
The deprojected knot velocities were calculated using the formula $V_f = \sqrt{| \bar{V}_r| ^2 + V_t^2}$.
The inclination angles of the knot ejections were estimated from the formula $I = \arctan (V_t / |\bar{V}_r|)$.

\begin{figure*}
\epsscale{1.0}
\plotone{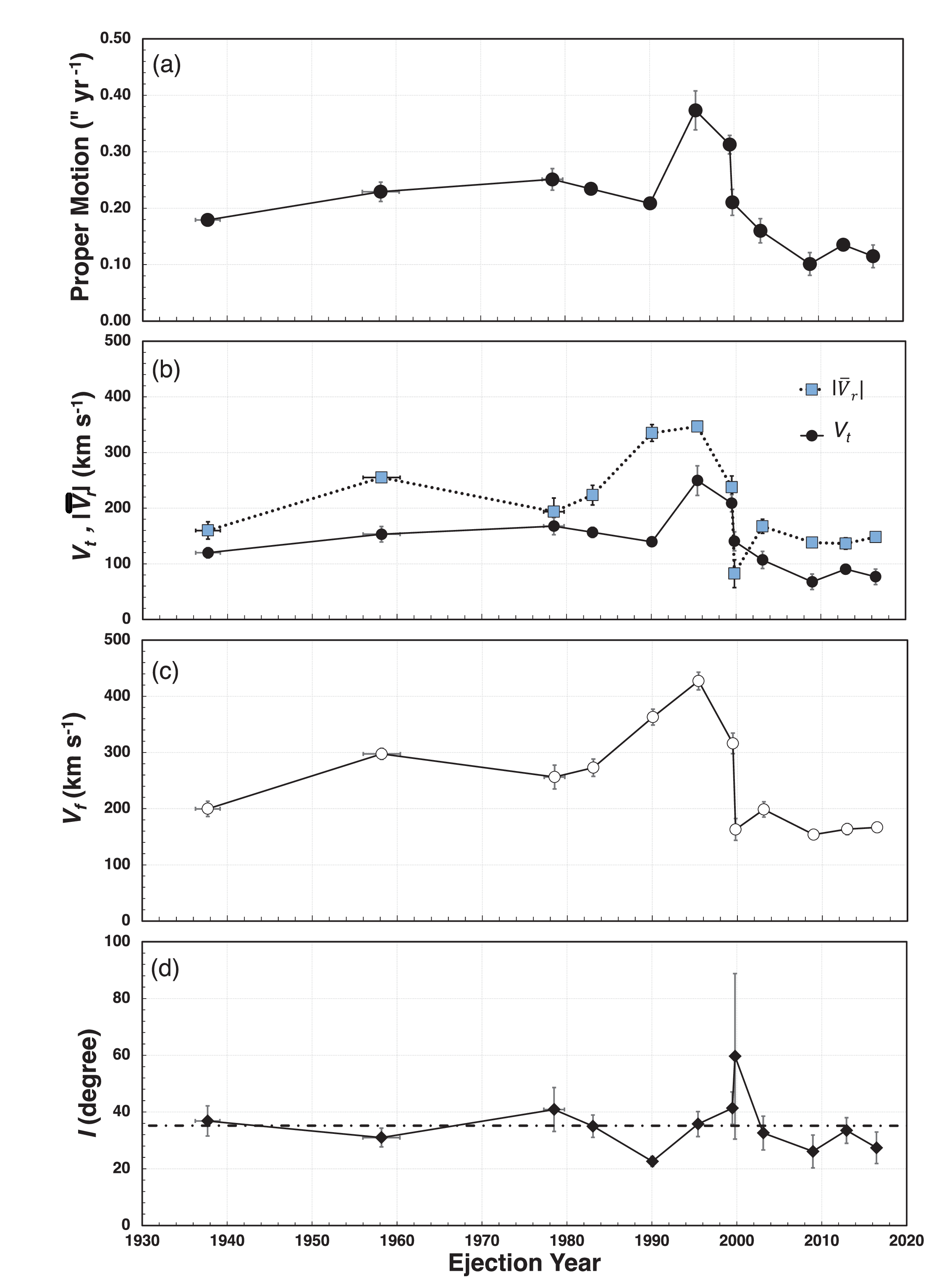}
\caption{Change of various parameters with the ejection date (yr) of knots. The parameters are: (a) Proper motions, (b) Radial ($|\bar{V}_r|$; $\blacksquare$ with dotted line) and tangential ($V_t$; $\bullet$ with solid line) velocities, (c) deprojected flow velocities ($V_f =\sqrt{|\bar{V}_r|^2 + V_t^2}$), and (d) Inclination angle ($I$). The dash-dotted line shows the average inclination angle of 35$\dotdeg$6.
\label{fig:EY_VI}}
\end{figure*}

Figure~\ref{fig:EY_VI}\textit{a} and \textit{b} show the proper motions and tangential velocities as functions of the ejection dates. 
The average proper motion for all the knot groups is 0\farcs21~$\pm$~0\farcs08 yr$^{-1}$, corresponding to a tangential velocity $V_t$ of 140~$\pm$~53 {\kms}. 
The proper motions remained approximately consistent with this average for the knots ejected from 1937 to 1990. 
However, there is a notable peak at 1995.44 associated with the knot A1, which has an elevated proper motion of 0\farcs37~$\pm$~0\farcs03 yr$^{-1}$, or a tangential velocity $V_t$ of 249~$\pm$~26 \kms. 
After 2003, the recorded proper motions have indicated slower movements less than 0\farcs15 yr$^{-1}$ ($V_t \sim$~100 \kms).

In Figure~\ref{fig:EY_VI}\textit{b}, the radial velocity ($|V_r|$)  is between 160 and 255~\kms\ for the knots ejected between 1937 and 1983.
There is then a sharp increase in 1990, reaching 335~$\pm$~15 \kms\ and peaking at 347~$\pm$~5 \kms\ in 1995.
By 2000, the radial velocity had significantly reduced, hovering around $\sim$~140 \kms\ from 2003 onwards, with knot kLVC marking a low point at 82~$\pm$~25 \kms.
The temporal variation in the radial velocity mirrors that of the tangential velocity, with both showing a pronounced rise around 1990 followed by a notable drop around 2000.
Figure~\ref{fig:EY_VI}\textit{c} displays a temporal variation in the deprojected knot velocity $V_f$.
It peaked at 427~$\pm$~16 \kms\ in 1995, reflecting the trends in radial and tangential velocities.
The radial velocities are typically $\sim$~1.5 times greater than the tangential velocities. 
The knot flow velocity $V_f$ was 200 to 300~km\,s$^{-1}$ between 1930 and 1983.
In the ejection dates of the 1990s, the velocity exceeded 300~km\,s$^{-1}$. 
However, it dropped to below 200~km\,s$^{-1}$ after 2000.
We will discuss the potential link between the rise in flow velocity during the 1990s and changes in the star-disk system activities. 
Such changes might have spurred rapid mass accretion onto the stellar surface while concurrently driving mass ejection from the star-disk system.
Figure~\ref{fig:EY_VI}\textit{d} shows the inclination angle ($I$).
It exhibits a roughly constant value of $\bar{I} =\overline{\arctan (V_t / |\bar{V}_r|)}$ =35$\dotdeg$6~$\pm$ 10$\dotdeg$9.

\begin{figure*}
\epsscale{1.1}
\plotone{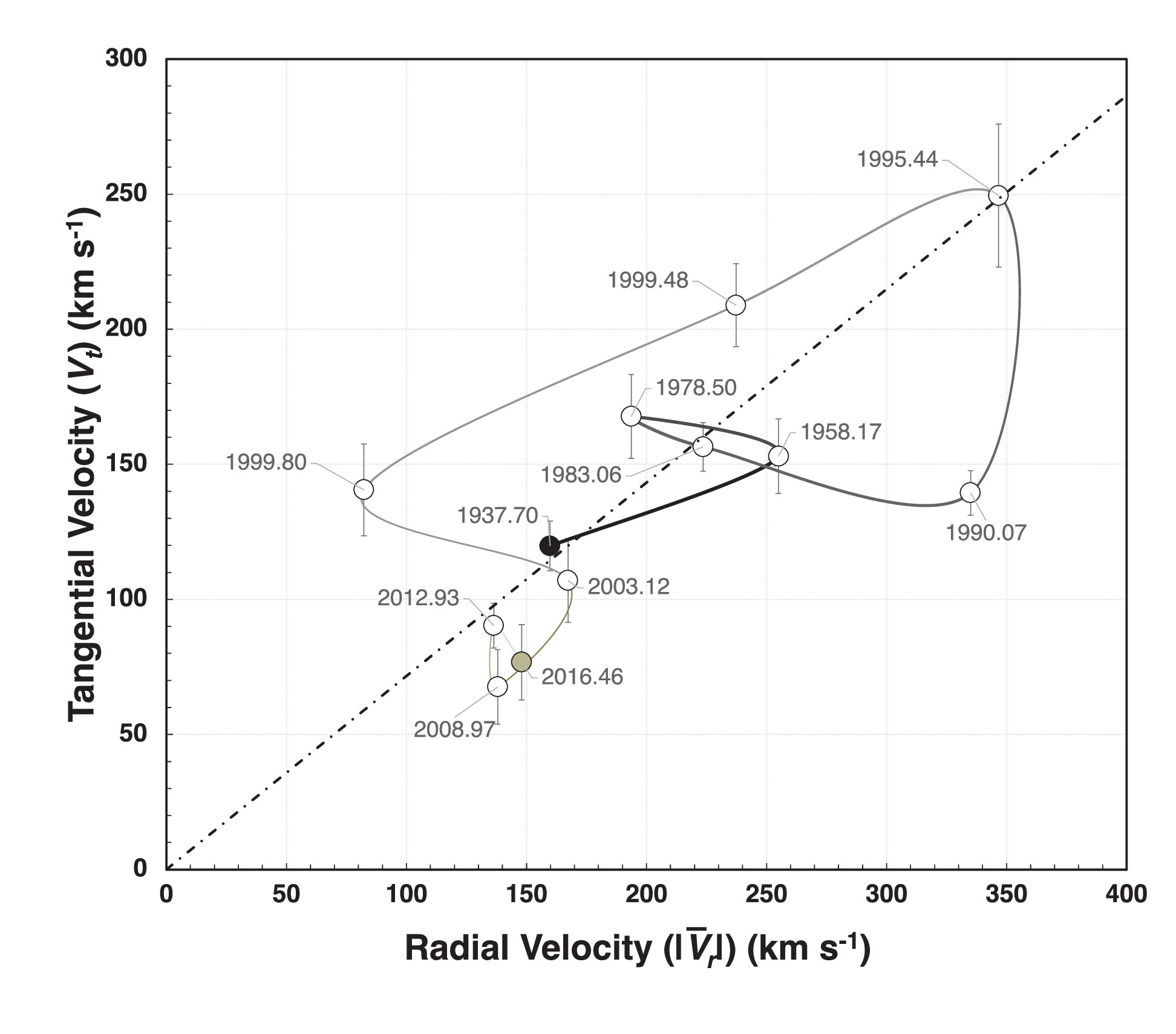}
\caption{Variation of knot velocities on the on the $| \bar{V}_r |$ vs. $V_t$ plane.
The earliest knot in 1937 is depicted in black, while the latest one in 2016 is shown in gray. Additionally, the connection line progressively thins out from 1937 to 2016.
The dash-dotted line indicates the average inclination angle of $\bar{I}~=$ 35$\dotdeg$6.
\label{fig:VrVt}}
\end{figure*}

In Figure~\ref{fig:VrVt}, the radial and tangential velocities of the knots are plotted and annotated with their respective ejection years for detailed examination of ejection angles.
The dash-dotted line indicates the angle of $\bar{I}$.
It appears that the direction of jet ejection varies over time.
Misaligned or wiggling shocked regions often occur due to variations in ejection angle over time \citep{Frank2014}.
In 1990, knot KHY1 was identified as the closest to the line of sight with an inclination angle of 22$\dotdeg$6~$\pm$~1$\dotdeg$8.
In 1999, knot kLVC registered an angle of 59$\dotdeg$6~$\pm$~29$\dotdeg$2. 
Since 2000, the inclination angle has shifted by $\sim$~6$\degree$ closer to the line of sight, averaging at 29$\dotdeg$6~$\pm$~7$\dotdeg$0, which is still within 1$\sigma$ of the mean inclination angle of 35$\dotdeg$6.
It is reasonable to assume that the direction of jet ejection varies across a certain range of angles.
This tendency can also be seen in the sky plane.
The early four knots (A, B, C, D) deviated $\pm$12$\arcdeg$ from the mean jet direction (PA~=~225$\arcdeg$) in proper motion vectors \citep{EM98}.
The jet wiggles about its mean direction, showing a bow shock or arc-shape feature at its tip \citep{Lavalley1997, Dougados2000, Takami2023}. 
The wiggle amounts to 0$\dotsec$2 deviated from the mean jet direction at $\sim$ 2$\dotsec$6 in 1997 \citep{Dougados2000} and $\sim$2$\arcsec$ in 2014 \citep{Takami2023}. 
The wiggling is corresponding to $\sim$4$-$6 degrees from the star on the sky plane. 
The ejection directions both on the sky plane and with respect to the line of sight vary within 1$\sigma$ level.
The irregular knot structure may be attributed to the variation in jet direction when the shock front in the knot is not fully resolved in space.

\section{Launching radii and mass accretion}\label{sec:accretion}

Numerous models propose that the velocity of a directly driven outflow, whether a jet, wind or another form, is roughly proportional to the Keplerian velocity at its launching radius \citep{Shu2000,Pudritz2007}. 
This suggests that high-velocity jets originate near the inner boundary of a disk, while low-velocity winds emerge from a slightly more outer region of the inner disk \citep{Pyo09, Frank2014}. 

\begin{figure*}
\plotone{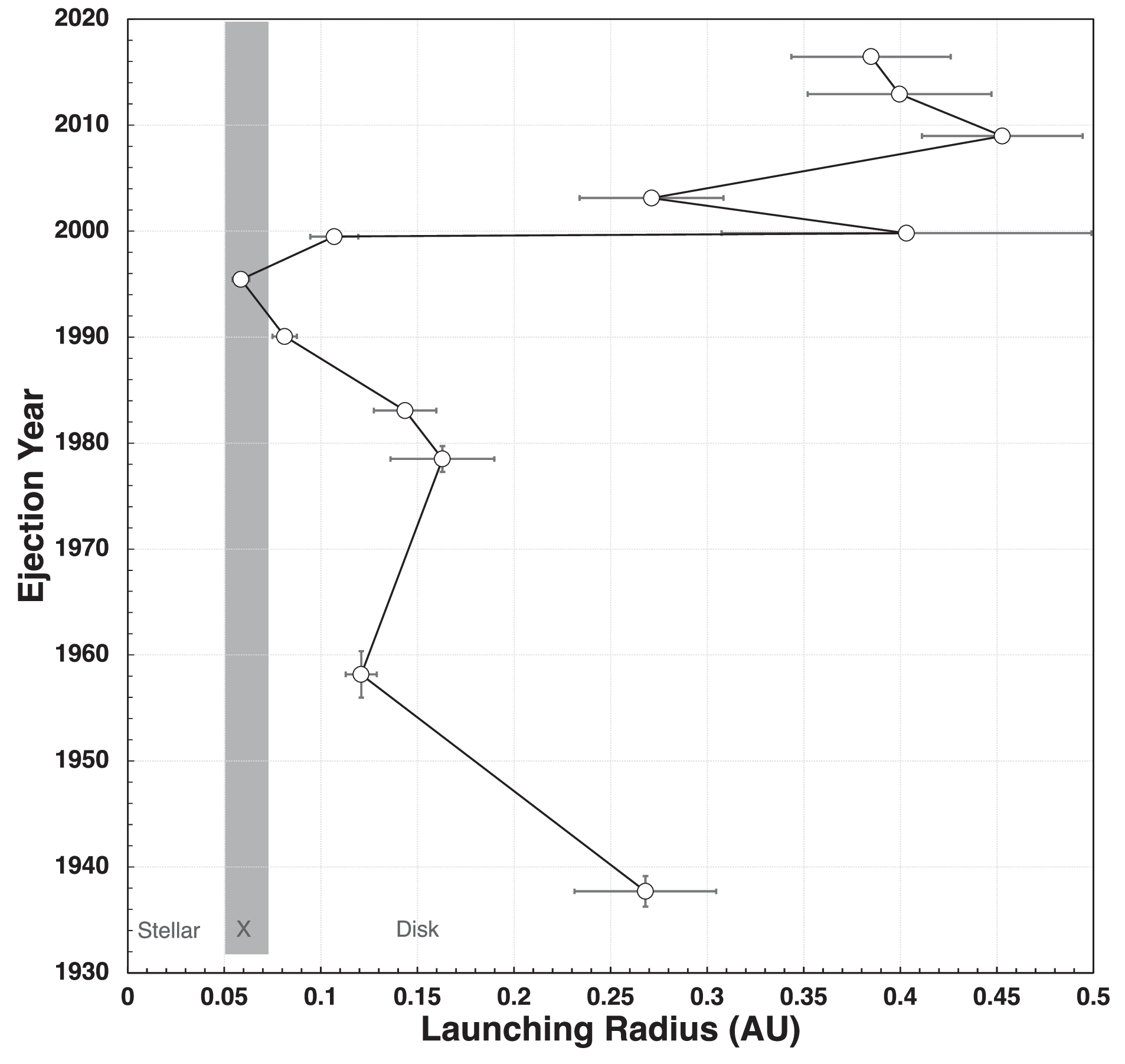}
\caption{
Variation of knot launching radii ($r_{\rm o}$; AU) from 1937 until 2016. The launching radii of the stationary MHD jet models are indicated: stellar winds($r_{\rm o} \la$ 0.05 AU), X-winds (gray-shaded area; 0.05$\la r_{\rm o} \la $0.07), and extended disc winds (0.07 $\la r_{\rm o}$) \citep{Ferreira2006}. The range of the X-winds launching radii is marked from 0.05 AU to 0.07 AU based on \citet{Shu1994,Shu1995,Shang1998}.    
\label{fig:launching_r}}    
\end{figure*}

To estimate the jet launching radius from $V_f$, we utilize the equation given by \citet{White2014_1}, operating under the assumption of a magneto-centrifugally driven wind. 

\begin{equation}
    (r_{\rm o}/0.1~AU) \simeq ~ \lbrack(109~{\rm km\,s}^{-1}/ {V}_f) ~ (r_A/r_{\rm o})\rbrack^2 ,
\end{equation} \label{eq:1}
where $r_A$ denotes the Alfv\'en radius, and $r_{\rm o}$ represents the launching radius.
We assumed parameters for the magnetic lever arm such that $(r_A / r_{\rm o})^2 = \lambda = $ 9. This is based on the scenario where the mass ejection rate from the outflow, $\dot{M}_{w}$, is approximately 10\% of the mass accretion rate, $\dot{M}_{a}$, i.e., $\dot{M}_{w} \simeq$ 0.1$\dot{M}_{a}$ \citep{Pudritz2007, KS2011, Eller2013}.
DG~Tau has $\dot{M}_{w} \sim 0.07 \dot{M}_{a}$ \citep{Coffey2008}.
We also used $M_*~=~$0.67 M$_{\sun}$ for the stellar mass of DG~Tau \citep{HEG95}.
The launching radii for the 12 identified knot groups, as detailed in Figure~\ref{fig:launching_r} and Table~\ref{tab:pm}, fall within a range of 0.06 to 0.45 AU. 
\citet{Coffey2007} reported similar radii (0.2-0.5 AU) for the higher velocity components estimated from the possible rotation signatures in the jet, assuming steady magneto-centrifugal acceleration.

Following the ejection of knot C in 1937 at a radius of 0.27~$\pm$~0.04 AU, $r_{\rm o}$ progressed inward by over 0.2 AU between 1958 and 1999. 
Knot B, ejected in 1958, had a launching point at $\sim$~0.12 AU with a velocity of 297 ~$\pm$~10 \kms. 
From the late 1970s onward, r$_{\rm o}$ neared the stellar surface, shifting from 0.16 AU to a close 0.06 AU by mid-1995. 
This radius equates to $\sim$~5.5 R$_*$ when considering the stellar radius of DG Tau as 2.3 R$_{\sun}$ \citep{HEG95}. 
This inward progression culminated with knot A1 achieving a maximum jet velocity of 427~$\pm$~16 \kms. 
From late 1999 to 2008, spanning approximately 10 years, $r_{\rm o}$ swiftly moved outward again, reaching 0.45 AU. 
Since 2008, however, it has decreased back to 0.35 AU. The variation in the launching radius suggests unsteady mass accretion for the last 80 years in the protoplanetary disk of DG Tau. 
The launching radii of the jets approached the stellar surface in 1995 and then expanded away, which may mean that the mass located in the radial range between 0.06 AU (5.5 $R_*$) and 0.45 AU accreted onto the star during the period.

\citet{Ferreira2006} delineate distinct regions from which extended disk winds, X-winds, and stellar winds originate. As illustrated in Figure~\ref{fig:launching_r}, these phenomena occupy specific radial zones: extended disk winds manifest beyond $r_{\rm o} \ga$ 0.07 AU, X-winds are situated between 0.05 AU $\la r_{\rm o} \la$ 0.07 AU, and stellar winds dominate the innermost region, existing at $r_{\rm o} \la$ 0.05 AU. The range of $r_{\rm o}$ associated with X-winds is informed by \citet{Shu1994,Shu1995} and \citet{Shang1998}. Notably, the A1 knot was launched at 0.06 AU, marking within the X-winds' launching domain. Subsequently, the launching radii retracted further from the star. This behavior might be linked to the efficient disruption of accretion and contraction in young stars by the phenomenon termed ``Reconnection X-wind" \citep{Ferreira2006}. 

From the fundamental energy and angular momentum conservation requirement, the following equation \citep[See][Eq. 4]{Coffey2015} is valuable in verifying consistency:
\begin{equation}
     \frac{\dot{M}_{w}}{\dot{M}_{a}} \leq \frac{1}{4(\lambda - 1)} \ln{\frac{r_{out}}{r_{in}}}
\end{equation}\label{eq:2} 
where $\lambda$ is the magnetic lever arm parameter ($= (r_A / r_{\rm o})^2$), $r_{out}$ is the maximum launching radius, $r_{in}$ is the minimum launching radius.
During 40 years, we can set $r_{out} =$ 0.45 AU and $r_{in} =$ 0.06 AU in this paper.
\citet{Agra2011} summarized that the range of the mass accretion rate ($\dot{M}_{a}$) is 1$-$5 $\times 10^{-7} M_{\sun}$ yr$^{-1}$ and the range of the mass ejection rate ($\dot{M}_{w}$) falls within 2.2$-$4.4 $\times 10^{-8} M_{\sun}$ yr$^{-1}$.
It gives the ratio of mass ejection and accretion rates of 0.04$-$0.44 for DG~Tau.
Applying these parameters implies that $\lambda \leq$ 2.1$-$12.4 is consistent with our assumed $\lambda \sim $ 9.

\section{Photometric variation and the flow velocities}

\begin{figure*}
\epsscale{0.9}
\plotone{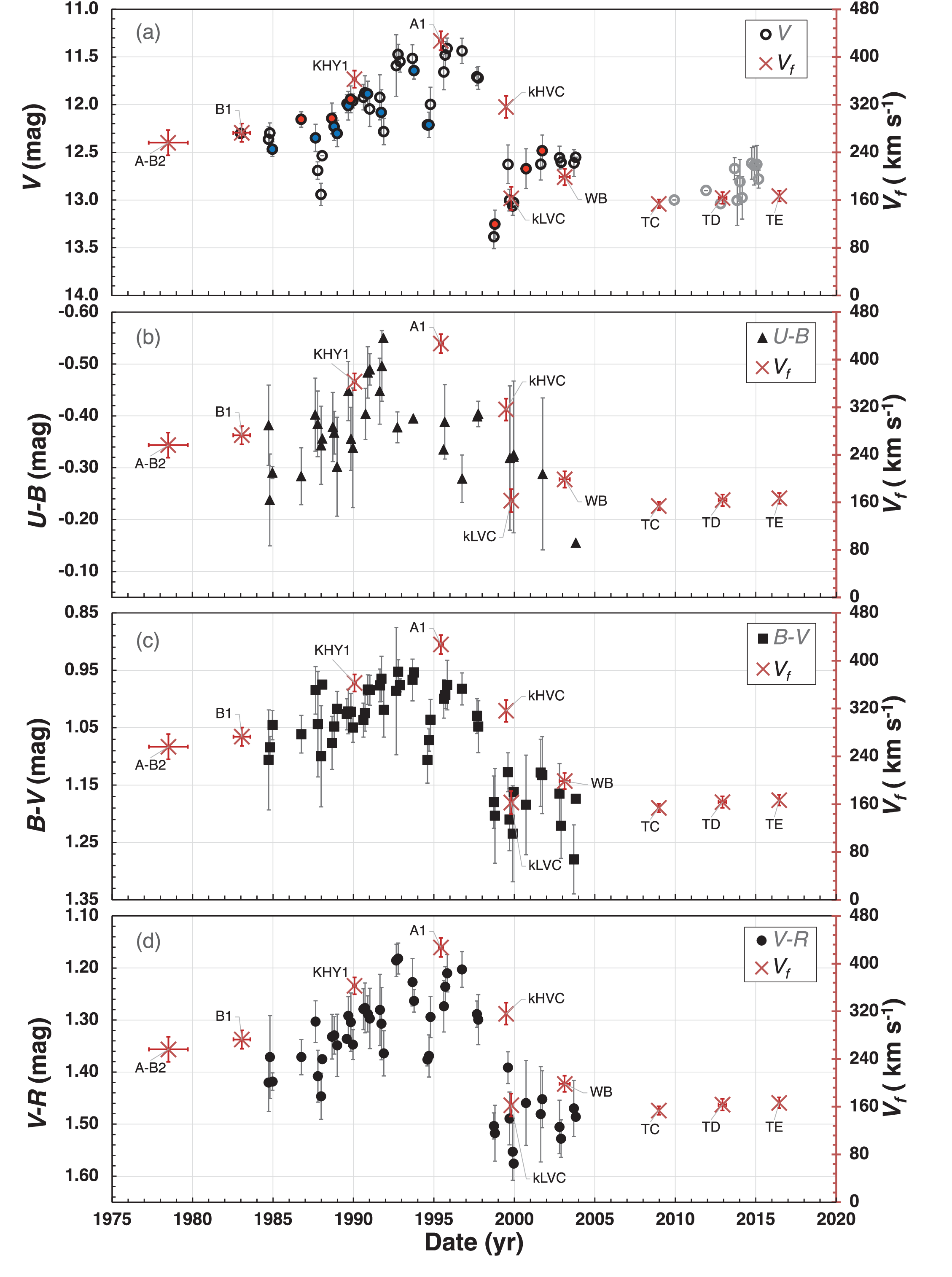}
\caption{
Comparison of DG Tau's (a) $V$, (b) $\ub$, (c) $\bv$, and (d) $\vr$ magnitudes (\textit{left axis})variation with deprojected knot velocities ($V_f$)(\textit{right axis}).
Opened circles, filled triangles, filled squares, and filled circles denote $V$, $\ub$, $\bv$, and $\vr$ magnitudes averaged over 10 rotational periods, 63 days, by \citet{Grankin2007} before 2005. The $V$ magnitudes after 2009 (gray color) are from the AAVSO database \citep{Kafka2015}.  
The circles filled with red and blue colors in the $V$ data represent points that are brighter and redder or fainter and bluer compared to the immediately preceding data points (See Figure~\ref{fig_dBV_dV}).
Crosses depict $V_f$ with the error bars showing standard deviations.  
\label{fig:V_Vf}
}
\end{figure*}

In Figure~\ref{fig:V_Vf}, we plotted the photometric variation of DG Tau in the $V$, $\ub$, $\bv$, and $\vr$ magnitude (left vertical axis) from 1983 to 2015 and the deprojected flow velocities ($V_f$, right vertical axis) at their respective ejection dates.
The photometric data from \citet{Grankin2007} from 1983 to 2003 is augmented by incorporating AAVSO $V$ magnitude observations from 2009 to 2015, predominantly undertaken by James McMath and Kenneth Menzies until March 2015 \citep{Kafka2015}.
We computed average magnitudes over ten rotational periods to exclusively discern long-term photometric variations and negate rotation-induced signals. 
 DG Tau's rotational period is 6.3 days, so each data point reflects an average magnitude spanning 63 days. 
 
The local maximum of the observed brightness rose from a magnitude of 12.1 to 11.1 in the $V$-band between 1985 and 1997, mirroring the rise in flow velocity. 
Between 1997 and 1998, it experienced a sharp decline, remaining dimmer than 12.4 in magnitude from 1998 to 2015.
Overall, the photometric variation in $V$-band aligns well with the changes in flow velocity.
A notable exception may be the kHVC in 1999, which followed the most pronounced dip in brightness in 1988.
The $\bv$, and $\vr$ color variations align also with the changes in the flow velocity while $\ub$ shows sharp decline (redder) in 1992 between the ejections of KHY1 and A1 knots.

\citet{Grankin2007} emphasized that DG Tau displays significant fluctuations in its average brightness over the years, beyond the abrupt seasonal variations. 
These pronounced fluctuations hint at considerable changes either in the accretion rate from the disk to the star or in the circumstellar extinction across multiple years. 
As previously discussed, a higher outflow velocity indicates that the outflow stems from regions nearer to the stellar surface.
The strong correlation between DG~Tau's brightness and outflow velocity, as depicted in Figure~\ref{fig:V_Vf}, indicates that the brightness surge in the 1990s is more likely due to an increase in the accretion rate rather than a reduction in circumstellar extinction.
Assuming constant mass density and gas volume, an increase in flow velocity suggests a corresponding increase in mass ejection rate. The higher mass ejection rate must be related to the higher mass accretion rate because they are proportional to each other as shown in the previous section, i.e. $\dot{M}_{w} / \dot{M}_{a} = 0.04 - 0.44$.

The rise in Br$\gamma$ emission-line luminosity during the 1990s further suggests an increase in the mass accretion rate. 
\citet{Muzerolle1998} demonstrated that the luminosities of Br$\gamma$ emission lines are closely correlated with the accretion luminosity ($L_{acc}$). 
They noted that in January 1998, when DG~Tau's brightness was around $V$ $\sim$ 11.7 mag, which was slightly dimmer than the brightest peak at $V$ $\sim$ 11.4 mag in October 1995, the equivalent width of the Br$\gamma$ emission, W(Br$\gamma$), was 14.1 \AA. 
This corresponds to a mass accretion rate of $\dot{M}_a =$ 7.9 $\times$ 10$^{-7}$ M$_\sun$ yr$^{-1}$. 
\citet{Coffey2008} estimated the mass ejection rate of 1.3 $\times$ 10$^{-7}$ M$_{\sun}$ yr$^{-1}$ with the data obtained by \citet{Bacciotti2000} in 1999 when knot A1 ejected in 1995, was located within 1$\arcsec$ from the star. 
For 1995$-$1999, the mass ejection and accretion rates marked the highest points. The ratio of $\dot{M}_{w} / \dot{M}_{a}$ was $\sim 0.2$.  
However, in October 2005, following a dimming of DG~Tau about one magnitude to $V$ $\sim$~12.5$-$13.0 mag, \citet{Agra2011} reported a reduced equivalent width of W(Br$\gamma$) = 5.6 \AA, indicating a lower mass accretion rate of $\dot{M}_a = $ 1 $\times$ 10$^{-7}$ M$_\sun$ yr$^{-1}$. 
Correspondingly, they estimated a decreased mass ejection rate of 3.3 $\pm$ 1.1 $\times$ 10$^{-8}$ M$_\sun$ yr$^{-1}$ using \FeII emission. 
These rates were notably smaller compared to those observed in 1998. 
The ratio $\dot{M}_{w} / \dot{M}_{a}$ was 0.2$-$0.4.  
\citet{Beck2010} observed a similar mass accretion rate, 9.6 $\times$ 10$^{-8}$ M$_\sun$ yr$^{-1}$, and estimated a lower limit for the mass ejection rate at 1.2 $\times$ 10$^{-8}$ M$_\sun$ yr$^{-1}$. Consequently, the ratio $\dot{M}_{w} / \dot{M}_{a}$ was inferred to be $>$ 0.13. 

\begin{figure*}
\epsscale{1.0}
\plotone{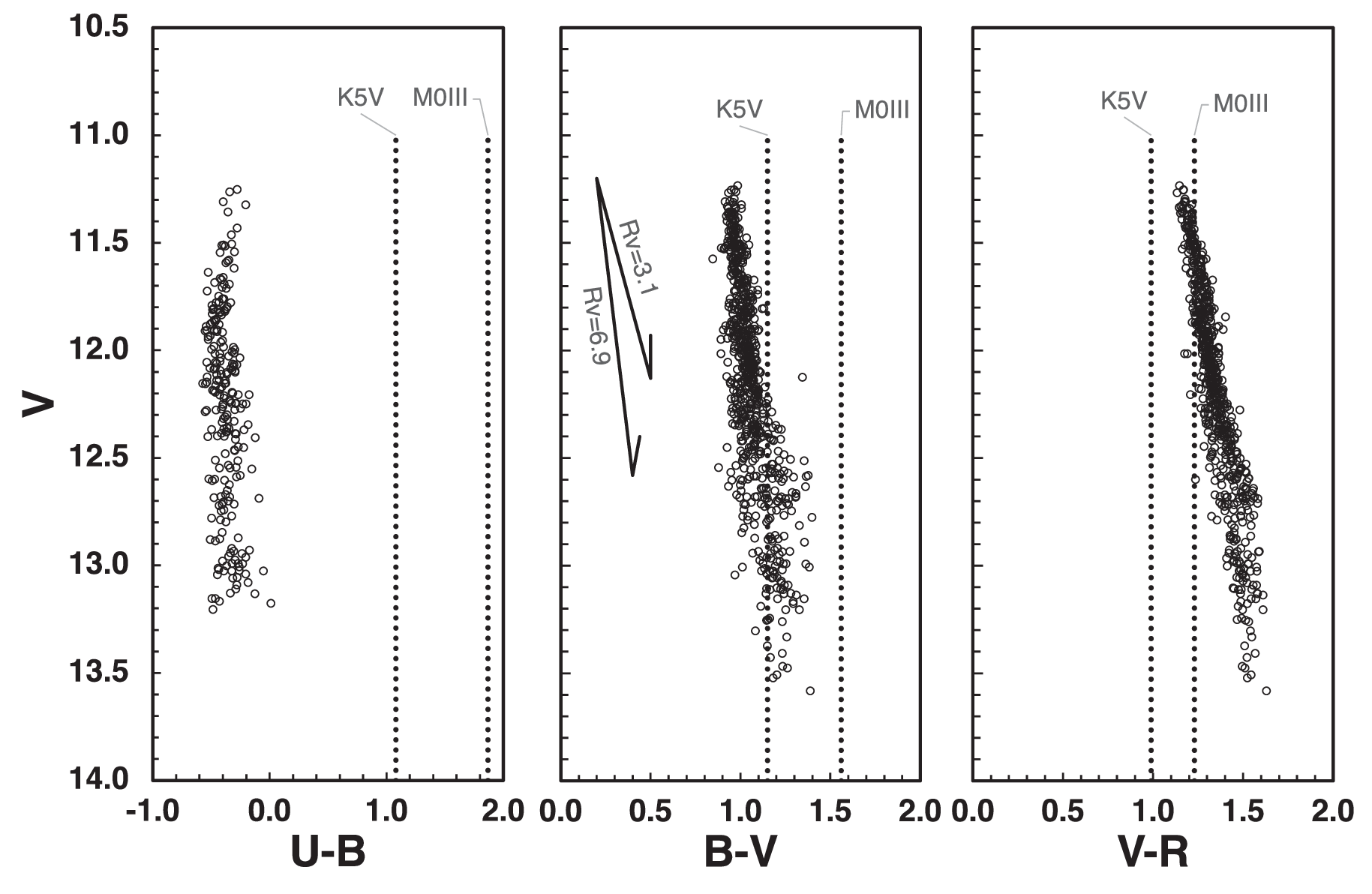}
\caption{Color-magnitude diagrams (CMD) of DG~Tau. (\textit{left}) \ub ~vs~$V$, (\textit{middle}) (\bv) vs $V$, (\textit{right}) \vr ~vs $V$. 
The data was taken from \citet{Grankin2007}. 
The vertical dotted lines in each panel indicate the intrinsic color of DG Tau for its spectral type of K5 \citep{HEG95} and M0 \citep{Herbst1994}. 
The luminosity classes of CTTSs are between giants (III) and dwarfs (V) \citep{Greene1996}. The arrows in the middle panel show the direction of extinction with $R_{\rm v} =$ 3.1 and 6.9 where  $R_{\rm v} =~A_{\rm V}$ / $E$(\bv).
\label{fig_cmd}
}
\end{figure*}

The variation in DG~Tau's photometric color further indicates that its increased brightness can be attributed to heightened accretion.
The color-magnitude diagrams depicted in Figure~\ref{fig_cmd} illustrate the variations of the colors $\ub$, $\bv$, and $\vr$ in relation to the photometric variation, based on the data from \citet{Grankin2007}.
The two dotted vertical lines represent the intrinsic color of DG Tau, falling between K5 and M0, as inferred from previously published spectral types: K5 \citep{HEG95}, K8.5 \citep{HP92}, and M0   \citep{Herbst1994}.
The observed $\ub$ color indices are noticeably bluer than the expected intrinsic color range.
Furthermore, the $\bv$ color indices are bluer than the intrinsic color when the star is at a brighter magnitude (11.2 $< V <$ 12.5), whereas they align with the intrinsic color when the star's brightness approaches its minimum magnitude (12.5 $< V <$ 13.7).
The large \ub\ excess in the left panel of Figure~\ref{fig_cmd} should be related to accretion from the inner edge of the disk onto the stellar surface because the $U$ band luminosity is proportional to the accretion luminosity \citep{Calvet2000}.
\citet{Vrba1993} showed that CTTSs typically have intrinsic colors of the normal photospheres at the minimum brightness and bluer colors at maximum brightness, regardless of whether the variability is periodic. 
They pointed out that accretion-driven hot spots are the predominant cause of increasing brightness. 

The color indices of \vr\ for DG~Tau are notably redder at minimum brightness. This pronounced redness in \vr\ can be ascribed to circumstellar extinction, analogous to the $V-I$ observed in the case of RW~Aur \citep{Petrov2007}.
Both the $\bv$ and \vr\ color-magnitude diagrams align best with a ratio of absolute to selective extinction $R_{\rm v}$ of approximately 6.9, as indicated by the arrow in the middle panel of Figure~\ref{fig_cmd}. 
This value surpasses the standard interstellar extinction value of $R_{\rm v} =$ 3.1. 
A larger $R_{\rm v}$ may suggest the presence of larger grain sizes \citep{Cardelli1989}. \citet{Isella2010} proposed that dust particles around DG Tau underwent reprocessing and grew in size, potentially reaching radii of at least 20 $\mu$m.

\begin{figure*}
\epsscale{1.0}
\plotone{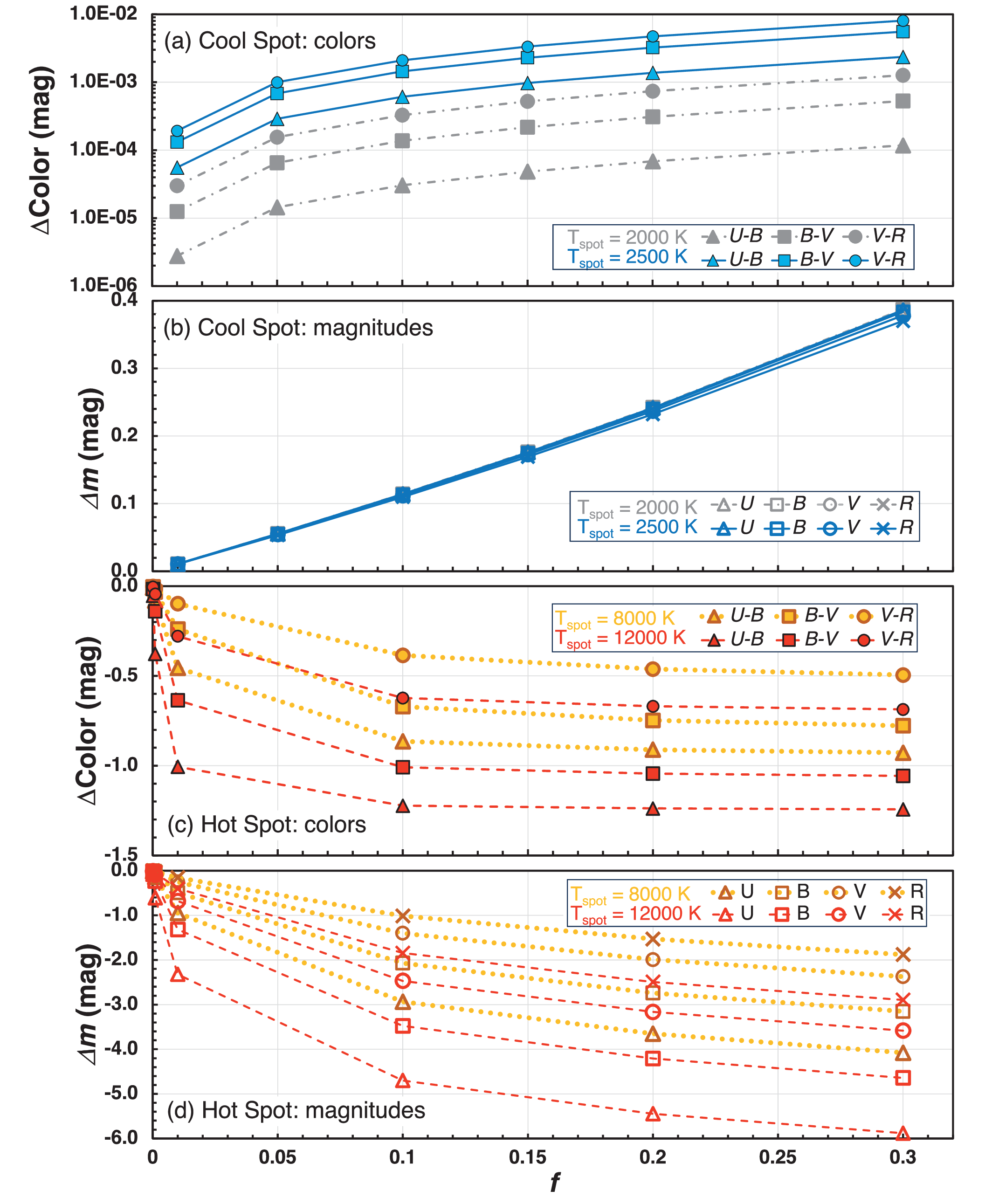}
\caption{The variations of the colors ($\Delta$($\ub$), $\Delta$($\bv$), $\Delta$($\vr$)) and magnitudes ($\Delta U$, $\Delta B$, $\Delta V$, $\Delta R$) by cool and hot spots model \citep{Carpenter2001, Vrba1993}. Panels (a) and (c) display color variations, while panels (b) and (d) depict magnitude variations resulting from cool and hot spots, respectively. 
The spots' coverage fractions ($f$) range from 0.01 to 0.3. 
The temperatures of the cool and hot spots are indicated by gray dash-dotted lines(2000 K), blue solid lines (2500 K), yellow dotted lines(8000 K), and red dashed lines (12000 K). 
The star's effective temperature is $T_{\ast}$ = 4000 K.
In panel (b), it is noteworthy that the $\Delta m$ of cool spots at $T_{\text{spot}} = 2000$ K and 2500 K exhibit minimal differences due to the narrow (500 K) temperature range.
\label{fig_spot_model}}
\end{figure*}

To quantitatively assess the hot and cold spots impact on the brightness, we calculated the magnitude difference, $\Delta m(\lambda)$, attributable to cool and hot spots, using formulas derived from  \citet{Carpenter2001} and \citet{Vrba1993}:  
\begin{equation}
    \Delta m(\lambda) = -2.5 \log \{1 - f[1.0 - B_{\lambda}(T_{spot})/B_{\lambda}(T_{\ast})]\}
\end{equation}\label{eq:3}
Here $f$ represents the fraction of the stellar photosphere covered by spots, $T_{spot}$ and $T_{\ast}$ denote the effective temperatures of the spot and star, respectively, and $B_{\lambda}(T)$ is the Planck function.
For DG Tau, which has a spectral type ranging from K5 to M0, we selected $T_{\ast} = 4000$ K, corresponding to temperatures between 3800 K and 4340 K \citep{HEG95, HP92, Herbst1994}.
Figure~\ref{fig_spot_model} illustrates the magnitude and color variations across the $UBVR$ bands. 
For cool spots, we consider $T_{spot}$ values of 2000 K and 2500 K. 
Meanwhile, for hot spots, $T_{spot}$ is taken as 8000 K and 12000 K. 
These variations are plotted for spot coverage fractions ($f$) of 0.01, 0.05, 0.1, 0.15, 0.2, and 0.3.

For the typical values of $T_{\ast} - T_{spot} \le $ 2000 K and $f \le$ 30 \% for cool spots, the magnitude and color variations are $\lesssim$ 0.4 mag and $\lesssim$ 0.009 mag, respectively.
As the coverage fraction of the cool spot increases, the star's magnitude becomes fainter and its color shifts to a redder value. 
With greater coverage of the stellar surface by cool spots, the shift in color becomes more pronounced at longer wavelengths: $\Delta(\ub) < \Delta(\bv) < \Delta(\vr)$.

For typical parameters, with $T_{spot} \leq 8000$ K and $f \leq 10\%$ for hot spots, the variations in magnitude and color are $\Delta V \lesssim$ 1.5 mag and $\Delta(\bv) \lesssim$ 0.7, respectively. 
More significant variations, such as $\Delta V \geq$ 3 mag and $\Delta(\bv) \geq$ 1 mag, can be expected if the hot spot has a high temperature ($T_{spot} = 12000$ K) and covers a more substantial portion of the stellar surface, with $f \geq 20\%$.
When hot spots maintain a consistent temperature, $T_{spot}$, and occupy a more extensive region on the stellar surface, the star appears brighter and exhibits a bluer color.
The variation in magnitude becomes more pronounced with decreasing wavelength: $\Delta R < \Delta V < \Delta B < \Delta U$. Similarly, color variations follow the trend: $\Delta$($\vr$) $<$ $\Delta$($\bv$) $<$ $\Delta$($\ub$).

\begin{figure*}
\epsscale{1.0}
\plotone{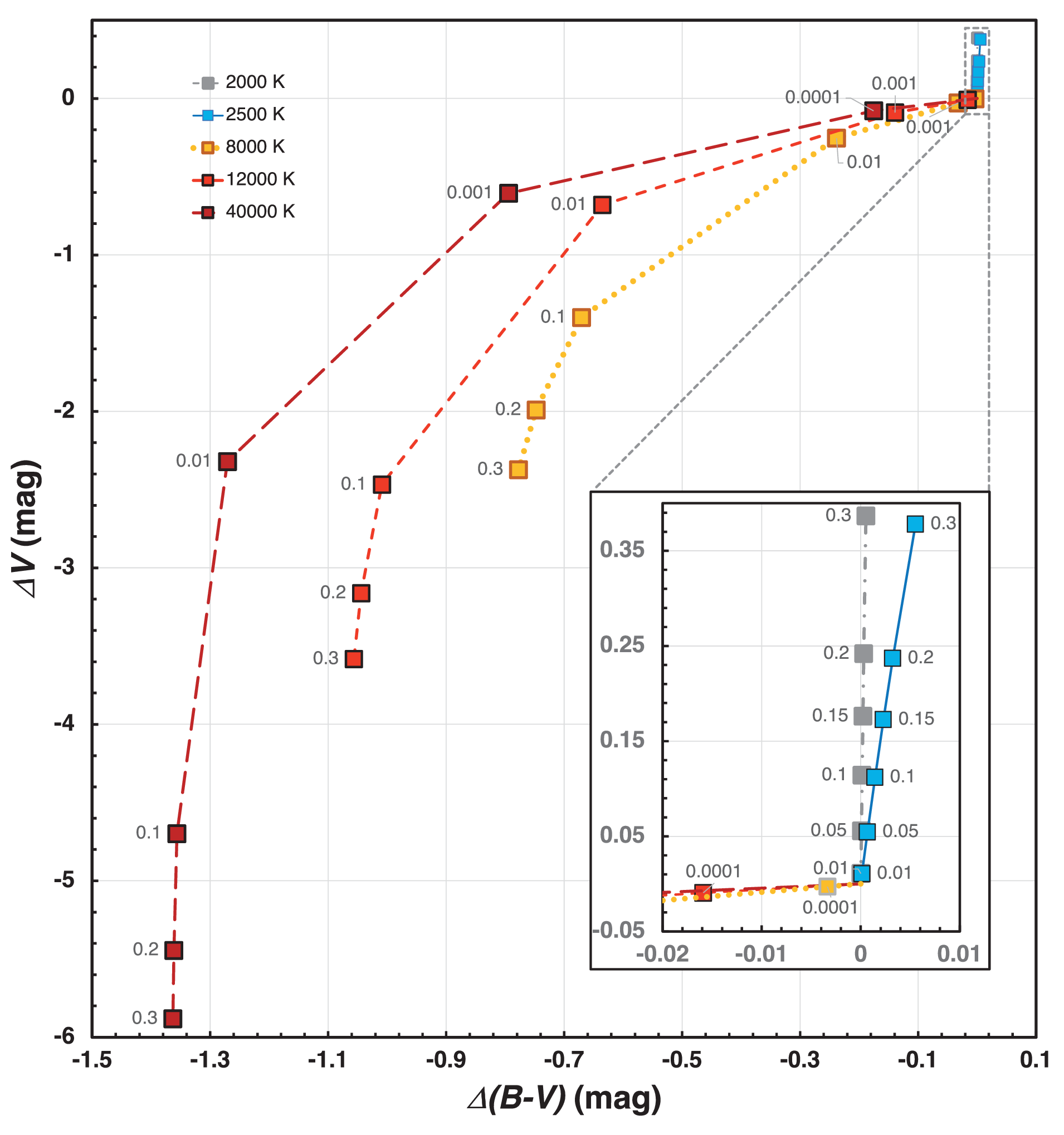}
\caption{Variation of $\Delta V$ and $\Delta$($\bv$) for cool and hot spot models \citep{Carpenter2001, Vrba1993}. 
The gray and blue filled squares in the upper right corner ($\Delta V \ge 0$ and $\Delta$($\bv$) $\ge 0$) represent cool spots with $T_{spot} =$ 2000 K (gray dash-dotted line) and 2500 K (blue solid line). 
The inset box provides a zoomed-in view of this region.
The yellow, red, and dark-red filled squares depict the hot spot model with $T_{spot} =$ 8000 K (yellow dotted line), 12000 K (red dashed line), and 40000 K (dark-red long-dashed line). 
The labels of the data points correspond to the coverage fraction ($f$) values: 0.0001, 0.001, 0.01, 0.1, 0.2, and 0.3.
\label{fig_spot_dcmd}}
\end{figure*}

Figure~\ref{fig_spot_dcmd} displays the variations $\Delta V$ and $\Delta$($\bv$) caused by both cool and hot spots. 
While both types of spots occupy the positive $\Delta V / \Delta ( \bv )$ space, they differ in their variation vector directions. 
From this figure, it is evident that the cool spot primarily accounts for the minor magnitude and color variation found in the upper right corner of the $\Delta V$ and $\Delta$($\bv$) space. 
As the value of $f$ increases, the rate of color change diminishes, even though the $V$ magnitude undergoes rapid alterations at a constant $T_{spot}$.
The dash-dotted lines link the data points across various $T_{spot}$ values but maintain a consistent $f$ value. 
Should $T_{spot}$ undergo changes without any alterations in spot coverage, the $\Delta V$ and $\Delta ( \bv )$ would shift along these dash-dotted lines.
We, therefore, deduce that the observed rise in the DG~Tau's $V$ magnitude in the 1990s can be attributed to an increase in the coverage fraction of hot spots, in line with the elevated accretion rate, rather than a decrease in the coverage fraction of cold spots.

\begin{figure*}
\epsscale{1.0}
\plotone{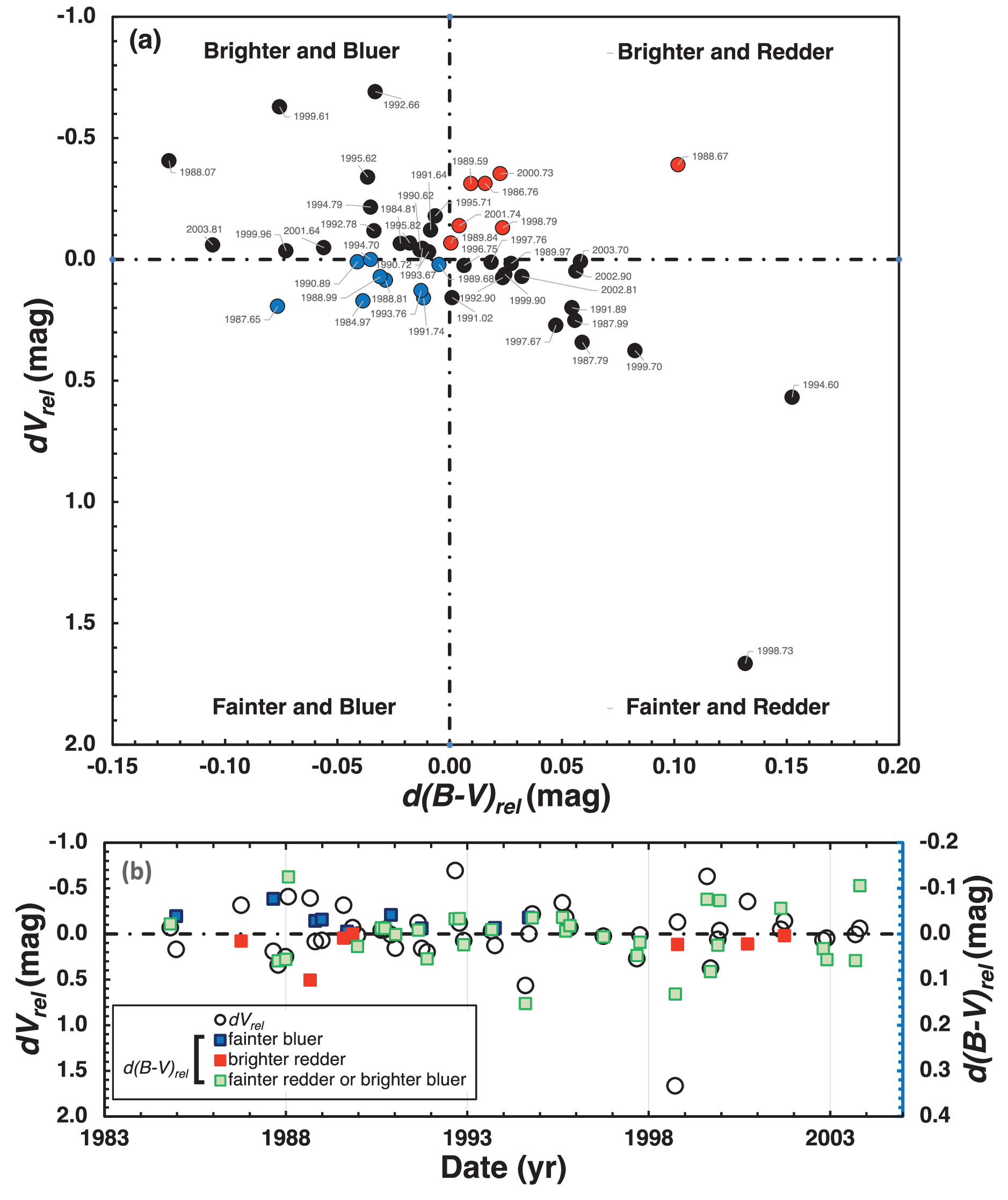}
\caption{Relative variations in color, $d(\bv)_{rel}$, and magnitude, $dV_{rel}$, are observed between successive data points. In the upper panel (a), changes in brightness and color corresponding to the four quadrants are classified: brighter \& bluer, brighter \& redder, fainter \& bluer, and fainter \& redder. The labels of the data points indicate the observed date.
In the lower panel (b), the relative variations along the observation dates are presented. Open circles represent $dV_{rel}$, while filled squares denote $d(\bv)_{rel}$. The filled squares are further classified based on relative changes: blue squares indicate fainter \& bluer, red squares indicate brighter \& redder, and green squares indicate fainter \& redder or brighter \& bluer.
Notably, all nine blue squares exist in the period from 1984 to 1995, during which the overall brightness of DG~Tau was increasing.
\label{fig_dBV_dV}
}
\end{figure*}

Figure~\ref{fig_dBV_dV}\textit{a} displays a plot comparing the relative magnitude $dV_{rel}$ with the color $d(\bv)_{rel}$, based on the averaged data points from Figure~\ref{fig:V_Vf}. 
These relative values are determined by the difference in magnitude and color relative to their preceding data points. 
Data positioned in the lower left and upper right quadrants yield a positive value for $dV_{rel} / d(\bv)_{rel}$, which suggests that they are either becoming brighter and bluer or fainter and redder compared to their preceding data points.

Increased disk mass accretion typically results in a trend toward brighter and bluer values, whereas a reduction in accretion leads to the converse trend, as pointed out by \citep{Carpenter2001}.
Variations in brightness, often spanning several magnitudes and exhibiting prolonged, irregular fluctuations, can arise from extinction. This effect occurs when a clump of circumstellar gas and dust obstructs the line of sight, making an object appear both fainter and redder.
Both scenarios—brightening due to mass accretion characterized by hot spots with a varying $f$ but consistent $T_{spot}$, and dimming from extinction effects—can explain the observed behavior in the two aforementioned quadrants.

Data in the lower right and upper left quadrants display a negative value for $dV_{rel} / d( \bv )_{rel}$, suggesting they either became brighter yet redder or dimmed while turning bluer relative to the previous measurements. This pattern does not align with typical extinction or mass accretion models.
For hot spots, such a trend can emerge if the temperature $T_{spot}$ decreases while its coverage fraction $f$ increases (e.g., transitioning from $T_{spot} =$ 12000 K to 8000 K, alongside an $f$ shift from 0.01 to 0.06, yields a brighter and redder outcome). 
Conversely, an increase in $T_{spot}$ coupled with a decrease in its coverage (e.g., moving from $T_{spot} =$ 8000 K to 12000 K, and $f$ from 0.06 to 0.01) produces a fainter, bluer look. 
Hence, the observed negative shift in $dV_{rel} / d( \bv )_{rel}$ underscores active mass accretion episodes marked by changes in both hot spot temperature and extent.
Figure~\ref{fig_dBV_dV}\textit{b} plots $dV_{rel}$ and $d( \bv )_{rel}$ along the observation dates. It shows all nine fainter and bluer data are located from 1984 to 1995 when the brightness of DG~Tau was generally increasing with some fluctuation. 

As discussed, the robust correlation between the variations in photometric magnitudes and flow velocities points to the predominant influence of mass accretion accompanied by hot spots. 
On the other hand, the observed reddening in the $\vr$ color for $V$ $\geq$ 12.5 suggests that DG~Tau underwent significant extinction after its sharp drop in brightness from 11.7 mag to 13.1 mag in the $V$ magnitude in 1998. 
This rapid dimming of V magnitude may be the result of the star being obscured by a dust barrier caused by the development of extensive disk winds.
As shown in Figure~~\ref{fig:V_Vf}, when knot kHVC was ejected at 1999.48, its brightness measured approximately $V \sim 12.6$ mag. This value is notably dimmer than the anticipated $V \sim 12$ mag, based on the apparent correlation observed in other knots between brightness and flow velocity.
At that time, both kHVC and kLVC were ejected nearly concurrently at 0.11~$\pm$~0.01 AU and 0.4~$\pm$~0.1 AU, respectively. 
\citet{Pyo03} postulated that the broad velocity range of kLVC was indicative of a disk wind originating over an extensive radius. 
Between 1990 and 2002, a low-velocity component with a broad velocity range was consistently detected in the PVDs of these knots \citep{SB93,Lavalley1997,Takami2002,Pyo03,Maurri2014}. 
Although these components are not counted among the 12 knots because of their diffuse emissions and undetermined proper motion, they may represent a pair, consisting of a high-velocity jet from the disk's inner edge and a disk wind covering a significant part of the outer disk.
This expansive disk wind might have led to the formation of a dust wall, inducing more pronounced extinction than anticipated, akin to the phenomenon detailed by \citet{Eller2014}.

\section{Summary and Conclusion}\label{sec:conclusion}

We have compiled an extensive dataset of the blueshifted jet knots from DG Tau spanning the last 40 years. 
By doing so, we were able to trace the mass ejection event back to 1937.
This dataset has been compared with the recorded photometric magnitude and color variations from 1983 to 2015.

The key findings can be summarized as follows:

\begin{enumerate}

\item Our analysis of \FeII\ data from 2007 does not reveal any distinct higher velocity component exceeding -200 \kms. This stands in stark contrast to the findings presented in \citet{Pyo03}. We found that a high-velocity flow, surpassing $|V_r|$ $>$ 200 \kms, was consistently detected from 1988 to 2005.  However, its intensity notably diminished after 2006. 

\item We categorized 51 observed knots spanning 40 years along DG~Tau's blueshifted jet, grouping them into 17 distinct knot groups. This classification was established under the assumption of ballistic motion, relying on consistent proper motions and comparable radial velocities. Our analysis focused on 12 moving knot groups, excluding five single-observation knots. Leveraging their proper motions and radial velocities, we derived estimations for ejection dates, inclination angles, deprojected flow velocities, and the launching radii of these knot groups.

\item We discovered a clear correlation between the deprojected flow velocities and the photometric magnitudes throughout the ejection dates. Initially, the deprojected flow velocity ranged between 200 and 300 \kms from 1937 to 1983. However, by the 1990s, the velocity surged beyond 300 km s$^{-1}$, peaking at over 420 \kms. This increase was followed by a steep decline to below 200 \kms between 1996 and 1999, maintaining lower velocities throughout the 2010s.
The variation in photometric magnitudes corresponded closely to these velocity changes from 1983 to 2015. The local maximum of the observed $V$ magnitude ascended from 12.1 to 11.1 between 1985 and 1997, mirroring the rise in the velocity during that period. However, between 1997 and 1998, a sharp decline in magnitude occurred, persisting dimmer than 12.4 from 1998 to 2015. Notably, the mass accretion rate around 1998 was approximately eight times higher than that in 2005, while the ratio of $\dot{M}_w / \dot{M}_a$ remained nearly constant at about 0.2. These variations in magnitude predominantly stem from active mass accretion, coupled with hot spot activities on the stellar surface.

\item The estimation of launching radii was derived from the deprojected velocity utilizing the magneto-centrifugal wind model. The launching radius of the jets varied between 0.06 and 0.45 AU. Across 58 years, spanning from 1937 to 1995, the launching radius gradually reduced from 0.27 AU to 0.06 AU, reaching the "X-winds" launching region by 1995. However, within a decade, by 2008, it swiftly retracted to a larger distance.

\item Between 1997 and 1999, when the kHVC ejection occurred, DG~Tau displayed a significantly fainter magnitude ($V$ $\sim$ 12.6 mag) compared to anticipated levels ($V$ $\sim$ 12.0 mag). This notable dimming likely resulted from substantial extinction caused by a dense dust wall, aligning with the sharp decrease in brightness observed during that specific time frame.

\item Several knots appear transiently, lasting only within a 2$\arcsec$ span. Thus, ejection events occur more frequently than the intervals between these knots indicate.

\end{enumerate}

\begin{acknowledgements}
We are grateful for the valuable input provided by the anonymous referee, which significantly contributed to enhancing the quality of this paper.
We express our sincere appreciation to the global community of observers for their contributions to the AAVSO International Database. 
The variable star observations they provided were instrumental in this study. 
Additionally, our research was enriched by the resources of NASA’s Astrophysics Data System and the SIMBAD database, maintained by CDS in Strasbourg, France.
\end{acknowledgements}


\clearpage
\begin{deluxetable*}{cclclclclclclcll}
\tabletypesize{\scriptsize} 
\tablecolumns{5}
\tablewidth{0pt}
\tablenum{1}
\tablecaption{Distances and radial velocities of knots \label{table:knots}}
\tablehead{
\colhead{Obs year} & \multicolumn{2}{c}{D} & \multicolumn{2}{c}{C} & \multicolumn{2}{c}{B} &\multicolumn{2}{c}{A-B2} & \multicolumn{2}{c}{B1}  & \multicolumn{2}{c}{B0} & \multicolumn{2}{c}{KHY1}  & \colhead{Reference} \\
\colhead{(yr)} & \colhead{($\arcsec$)} & \colhead{(\kms)} & \colhead{($\arcsec$)} & \colhead{(\kms)} & \colhead{($\arcsec$)} & \colhead{(\kms)} & \colhead{($\arcsec$)} & \colhead{(\kms)} & \colhead{($\arcsec$)} & \colhead{(\kms)} & \colhead{($\arcsec$)} & \colhead{(\kms)} & \colhead{($\arcsec$)} & \colhead{(\kms)} &\colhead{}
}
\startdata
1984.77 & \nodata & \nodata & 8.50    & -136    & \nodata & \nodata & \nodata & \nodata & \nodata & \nodata       & \nodata & \nodata   & \nodata & \nodata       &  1 \\
1986.99 & 10.70   & \nodata & 8.70    & -156    & 6.60    & -255    & 2.50    & -161    & \nodata & \nodata       & \nodata & \nodata   & \nodata & \nodata       &  2 \\
1991.23 & \nodata & \nodata & \nodata & \nodata & \nodata & \nodata & \nodata & \nodata & \nodata & \nodata       & \nodata & \nodata   & 0.25    & \nodata       &  3\\
1992.62 & \nodata & \nodata & \nodata & \nodata & \nodata & \nodata & 3.25    & -200    & 2.25    & -255          & \nodata & \nodata   & 0.52    & -350 $^{(a)}$ &  4 \\
1994.84 & \nodata & \nodata & \nodata & \nodata & \nodata & \nodata & 3.90    & \nodata & 2.70    & -210          & \nodata & \nodata   & 1.00    & -320 $^{(b)}$ &  5\\
1997.04 & \nodata & \nodata & \nodata & \nodata & \nodata & \nodata & \nodata & \nodata & 3.30    & -235          & \nodata & \nodata   & \nodata & \nodata       &  6 \\
1997.99 & \nodata & \nodata & \nodata & \nodata & \nodata & \nodata & \nodata & \nodata & 3.40    & \nodata       & \nodata & \nodata   & \nodata & \nodata       &  6\\
1998.07 & \nodata & \nodata & \nodata & \nodata & \nodata & \nodata & \nodata & \nodata & 3.60    & -235          & \nodata & \nodata   & \nodata & \nodata       &  7\\
1998.10 & \nodata & \nodata & 10.65   & \nodata & \nodata & \nodata & \nodata & \nodata & \nodata & \nodata       & \nodata & \nodata   & \nodata & \nodata       &  8\\
1999.04 & \nodata & \nodata & \nodata & \nodata & \nodata & \nodata & \nodata & \nodata & 3.80    & -210          & 3.30    & -255      & \nodata & \nodata       & 9, 10\\
2002.71 & \nodata & \nodata & 12.00   & -175    & \nodata & \nodata & \nodata & \nodata & 4.60    & -220 $^{(c)}$ & \nodata & \nodata   & \nodata & \nodata       & 13 \\
2012.87 & \nodata & \nodata & 13.70   & -172    & \nodata & \nodata & 8.75    & -220    & 6.74    & -200          & \nodata & \nodata   & \nodata & \nodata       & 8\\
2014.89 & \nodata & \nodata & 13.50   & \nodata & \nodata & \nodata & \nodata & \nodata & 7.65    & \nodata       & \nodata & \nodata   & \nodata & \nodata       & 8 \\
\enddata
\tablecomments{(a) knot C in the \citet{SB93}, (b) an extension of the HVB of \citet{Lavalley1997}, (c) knot B in the \citet{Whelan2004}.}
\tablerefs{(1) \citet{Mundt1987}, (2) \citet{EM98}, (3) \citet{Kepner1993}, (4) \citet{SB93}, (5) \citet{Lavalley1997}, (6) \citet{Dougados2000}, (7) \citet{Lavalley2000}, (8) \citet{Oh2015}, (9)\citet{Bacciotti2000}, (10) \citet{Maurri2014}, (11) \citet{Takami2002}, (12) \citet{Pyo03},  (13) \citet{Whelan2004}, (14) \citet{White2014_1}, (15)\citet{Agra2011}, (16) This paper, (17) \citet{Takami2023}}
\end{deluxetable*}

\begin{deluxetable*}{cclclclcrclclcll}
\tabletypesize{\scriptsize} 
\tablecolumns{5}
\tablewidth{0pt}
\tablenum{1}
\tablecaption{Continued \label{table:knots_2}}
\tablehead{
\colhead{Obs year} & \multicolumn{2}{c}{A1} & \multicolumn{2}{c}{A2} &\multicolumn{2}{c}{kHVC} & \multicolumn{2}{c}{kLVC} & \multicolumn{2}{c}{WC} &\multicolumn{2}{c}{WB} & \multicolumn{2}{c}{PB} & \colhead{Reference} \\
\colhead{(yr)} & \colhead{($\arcsec$)} & \colhead{(\kms)} & \colhead{($\arcsec$)} & \colhead{(\kms)} & \colhead{($\arcsec$)} & \colhead{(\kms)} & \colhead{($\arcsec$)} & \colhead{(\kms)} & \colhead{($\arcsec$)} & \colhead{(\kms)} & \colhead{($\arcsec$)} & \colhead{(\kms)} & \colhead{($\arcsec$)} & \colhead{(\kms)} & \colhead{}
}
\startdata
1997.04 & 0.60    & -350    & \nodata & \nodata & \nodata & \nodata    & \nodata  & \nodata   & \nodata & \nodata       & \nodata & \nodata        & \nodata & \nodata    & 6 \\
1997.99 & 1.00    & \nodata & \nodata & \nodata & \nodata & \nodata    & \nodata  & \nodata   & \nodata & \nodata       & \nodata & \nodata        & \nodata & \nodata    & 6 \\
1998.07 & 0.93    & -350    & \nodata & \nodata & \nodata & \nodata    & \nodata  & \nodata   & \nodata & \nodata       & \nodata & \nodata        & \nodata & \nodata    & 7 \\
1999.04 & 1.35    & -340    & 0.45    & -230    & \nodata & \nodata    & \nodata  & \nodata   & \nodata & \nodata       & \nodata & \nodata        & \nodata & \nodata    & 9,10\\
2000.95 & \nodata & \nodata & \nodata & \nodata & 0.45    & -220       & 0.23     & -100      & \nodata & \nodata       & \nodata & \nodata        & \nodata & \nodata    & 11\\
2001.83 & \nodata & \nodata & \nodata & \nodata & 0.75    & -226       & 0.45     & -100      & \nodata & \nodata       & \nodata & \nodata        & \nodata & \nodata    & 12\\
2002.71 & \nodata & \nodata & \nodata & \nodata & 1.0     & -266       & 0.60     & -47       & \nodata & \nodata       & \nodata & \nodata        & \nodata & \nodata    & 13 \\
2005.87 & \nodata & \nodata & \nodata & \nodata & \nodata & \nodata    & \nodata  & \nodata   & 1.24    & -250 $^{(d)}$ & 0.40    & -180 $^{(e)}$  & \nodata & \nodata    & 14, 15\\
2006.98 & \nodata & \nodata & \nodata & \nodata & \nodata & \nodata    & \nodata  & \nodata   & \nodata & \nodata       & 0.61    & \nodata       & \nodata  & \nodata    & 14\\
2007.12 & \nodata & \nodata & \nodata & \nodata & \nodata & \nodata    & \nodata  & \nodata   & \nodata & \nodata       & 0.72    & -154 $^{(f)}$ & 0.30     & -154       & 16\\
2009.88 & \nodata & \nodata & \nodata & \nodata & \nodata & \nodata    & \nodata  & \nodata   & \nodata & \nodata       & 1.07    & \nodata       & \nodata  & \nodata    & 14\\
\enddata
\tablecomments{(d) and (e) knots A4 and A5 of \citet{Agra2011} respectively, (f) knot PC of this paper.}
\end{deluxetable*}

\begin{deluxetable*}{cclclcll}
\tabletypesize{\scriptsize} 
\tablecolumns{5}
\tablewidth{0pt}
\tablenum{1}
\tablecaption{Continued \label{table:knots_3}}
\tablehead{
\colhead{Obs year} & \multicolumn{2}{c}{TC} & \multicolumn{2}{c}{TD} & \multicolumn{2}{c}{TE} & \colhead{Reference} \\
\colhead{(yr)} & \colhead{($\arcsec$)} & \colhead{(\kms)} & \colhead{($\arcsec$)} & \colhead{(\kms)} & \colhead{($\arcsec$)} & \colhead{(\kms)} &\colhead{}
}
\startdata
2014.16 & 0.53    & -142      & \nodata  & \nodata & \nodata & \nodata    & 17\\
2014.99 & 0.61    & -134      & 0.31     & -120    & \nodata & \nodata    & 17 \\
2017.13 & \nodata & \nodata   & 0.51     & -152    & \nodata & \nodata    & 17\\
2017.98 & \nodata & \nodata   & 0.69     & -134    & \nodata & \nodata    & 17\\
2018.91 & \nodata & \nodata   & 0.82     & -136    & 0.28    & -141       & 17\\
2019.79 & \nodata & \nodata   & 0.94     & -140    & 0.38    & -155       & 17 \\
\enddata
\end{deluxetable*}

\clearpage
\begin{deluxetable*}{clllllcl}
\tabletypesize{\scriptsize} 
\tablecolumns{5}
\tablewidth{0pt}
\tablenum{2}
\tablecaption{Proper motions, velocities, inclination angles, and launching radii \label{tab:pm}}
\tablehead{
\colhead{Knot Group}  & \colhead{Proper motion} & \colhead{$|\bar{V}_r|$  $^{(a)}$} & \colhead{$V_t$ $^{(b)}$} & \colhead{$V_f$ $^{(c)}$} & \colhead{$I$ $^{(d)}$}     & \colhead{Launching radius ($r_o$)} & \colhead{Ejection Date }  \\
\colhead{}     & \colhead{($\arcsec$ yr$^{-1}$)} & \colhead{(km s$^{-1}$)} & \colhead{(km s$^{-1}$)} & \colhead{(km s$^{-1}$)} & \colhead{($\degree$)} & \colhead{(AU)} & \colhead{(yr)} 
}
\startdata
    C     & 0.18~$\pm$~0.01 & 160~$\pm$~16 & 120~$\pm$~9 & 200~$\pm$~14 & 36.9~$\pm$~5.3 & 0.27~$\pm$~0.04 & 1937.70~$\pm$~1.45 \\
    B     & 0.23~$\pm$~0.02{$^{(e)}$} & 255~$\pm$~8 & 153~$\pm$~14 & 297~$\pm$~10 & 31.0~$\pm$~3.3 & 0.12~$\pm$~0.01 & 1958.17~$\pm$~2.18 \\
    A-B2  & 0.25~$\pm$~0.02 & 194~$\pm$~24 & 168~$\pm$~16 & 256~$\pm$~21 & 40.9~$\pm$~7.7 & 0.16~$\pm$~0.03 & 1978.50~$\pm$~1.21  \\
    B1    & 0.23~$\pm$~0.01 & 224~$\pm$~18 & 156~$\pm$~9 & 273~$\pm$~15 & 35.0~$\pm$~3.9 & 0.14~$\pm$~0.02 & 1983.06~$\pm$~0.52 \\
    KHY1  & 0.21~$\pm$~0.01 & 335~$\pm$~15 & 139~$\pm$~8 & 363~$\pm$~14 & 22.6~$\pm$~1.8 & 0.08~$\pm$~0.01 & 1990.07~$\pm$~0.11 \\
    A1    & 0.37~$\pm$~0.03 & 347~$\pm$~5  & 249~$\pm$~26 & 427~$\pm$~16 & 35.7~$\pm$~4.4 & 0.06~$\pm$~0.00 & 1995.44~$\pm$~0.13 \\
    kHVC   & 0.31~$\pm$~0.02 & 237~$\pm$~20 & 209~$\pm$~15 & 316~$\pm$~18 & 41.4~$\pm$~5.7 & 0.11~$\pm$~0.01 & 1999.48~$\pm$~0.05 \\
    kLVC   & 0.21~$\pm$~0.02 &  82~$\pm$~25 & 141~$\pm$~17 & 163~$\pm$~19 & 59.6~$\pm$~29.2 & 0.40~$\pm$~0.10 & 1999.80~$\pm$~0.10 \\
    WB    & 0.16~$\pm$~0.02 & 167~$\pm$~13 & 108~$\pm$~15 & 199~$\pm$~14 & 32.8~$\pm$~6.0 & 0.27~$\pm$~0.04 & 2003.12~$\pm$~0.32 \\
    TC    & 0.10~$\pm$~0.02 & 138~$\pm$~4  & 68~$\pm$~14 & 154~$\pm$~7 & 26.1~$\pm$~5.8 & 0.45~$\pm$~0.04 & 2008.97~$\pm$~0.02 \\
    TD    & 0.14~$\pm$~0.01 & 136~$\pm$~10 & 90~$\pm$~8 & 164~$\pm$~10 & 33.5~$\pm$~4.5 & 0.40~$\pm$~0.05 & 2012.93~$\pm$~0.24 \\
    TE    & 0.11~$\pm$~0.02 & 148~$\pm$~7  & 77~$\pm$~14 & 167~$\pm$~9 & 27.4~$\pm$~5.6 & 0.38~$\pm$~0.04 & 2016.46~$\pm$~0.02 \\
    \hline
    Average & 0.21~$\pm$~0.08 & 207~$\pm$~74 & 140~$\pm$~53 & 248~$\pm$~90 & 35.6~$\pm$~10.9 & 0.24~$\pm$~0.15  &    \nodata   \\
\enddata
\tablecomments{
(a) Average radial velocity for each knot group.
(b) Tangential velocity for each knot group calculated under the assumed distance of 141~$\pm$~7 pc to TMC, taken from the GAIA DR2 edition \citep{Zucker2019}.
(c) deprojected flow velocity $V_f = \sqrt{{\bar{V}_r}^2 + {V_t}^2}$.
(d) Inclination angle with respect to the line of sight, $I~= \arctan(V_t / | \bar{V}_r |)$.
(e) from \citet{EM98}.}
\end{deluxetable*}

\clearpage

\clearpage

\clearpage

\clearpage

\clearpage

\clearpage

\clearpage

\clearpage

\clearpage

\clearpage

\clearpage


\begin{thebibliography}{} 
\bibitem[Agra-Amboage et al.(2011)]{Agra2011} Agra-Amboage, V., Dougados, C., Cabrit, S., \& Reunanen, J.\ 2011, \aap, 532, AA59 
\bibitem[Babina, Artemenko, \& Petrov(2016)]{Babina2016} Babina, E.~V., Artemenko, S.~A., \& Petrov, R.~P.\ 2016, Astronomy Letters, 42, 193. 
\bibitem[Bacciotti et al.(2000)]{Bacciotti2000} Bacciotti, F., Mundt, R., Ray, T.~P., et al.\ 2000, \apjl, 537, L49 
\bibitem[Beck et al.(2008)]{Beck2008} Beck, T.~L., McGregor, P.~J., Takami, M., \& Pyo, T.-S.\ 2008, \apj, 676, 472 
\bibitem[Beck, Bary, \& McGregor(2010)]{Beck2010} Beck, T.~L., Bary, J.~S., \& McGregor, P.~J.\ 2010, \apj, 722, 1360
\bibitem[Calvet et al.(2000)]{Calvet2000} Calvet, N., Hartmann, L., \& Strom, S.~E.\ 2000, in Mannings, V., Boss, A., Russell, S. Eds., Protostars and Planets IV, University of Arizona Press, Tucson, pp 377-398
\bibitem[Cardelli, Clayton, \& Mathis(1989)]{Cardelli1989} Cardelli, J.~A., Clayton, G.~C., \& Mathis, J.~S. 1989, \apj, 345, 245
\bibitem[Carpenter et al.(2001)]{Carpenter2001} Carpenter, J.~M., Hillenbrand, L.~A., \& Skrutskie, M.~F.\ 2001, \aj, 121, 3160 
\bibitem[Chou et al.(2013)]{Chou2013} Chou, M.-Y., Takami, M., Manset, N., et al.\ 2013, \aj, 145, 108 
\bibitem[Coffey et al.(2007)]{Coffey2007} Coffey, D., Bacciotti, F., Ray, T.~P., Eisl{\"o}ffel, J., Woitas, J.\ 2007, \apj, 663, 350
\bibitem[Coffey, Bacciotti, \& Podio(2008)]{Coffey2008} Coffey, D., Bacciotti, F., Podio, L.\ 2008, \apj, 689, 1112
\bibitem[Coffey et al.(2015)]{Coffey2015} Coffey, D., Dougados, C., Cabrit, S., Pety, J., \& Bacciotti, F.\ 2015, \apj, 804, 2
\bibitem[Dougados et al.(2000)]{Dougados2000} Dougados, C., Cabrit, S., Lavalley, C., \& M{\'e}nard, F.\ 2000, \aap, 357, L61 
\bibitem[Eisl{\"o}ffel \& Mundt(1998)]{EM98} Eisl{\"o}ffel, J., \& Mundt, R.\ 1998, \aj, 115, 1554 
\bibitem[Ellerbroek et al.(2013)]{Eller2013} Ellerbroek, L.~E., Podio, L., Kaper, L., Sana, H., Huppenkothen, D., de Koter, A., Monaco, L. 2013, \aap, 551,A5
\bibitem[Ellerbroek et al.(2014)]{Eller2014} Ellerbroek, L. E., Podio, L., Dougados, C., et al.\ 2014, \aap, 563, 87
\bibitem[Ferreira, Dougados, \& Cabrit(2006)]{Ferreira2006} Ferreira, J., Dougados, C., \& Cabrit, S.\ 2006, \aap, 453, 785
\bibitem[Frank et al.(2014)]{Frank2014} Frank, A., Ray, T. P., Cabrit, S., et al. 2014, in Protostars and Planets VI, ed.
H. Beuther et al. (Tucson, AZ: Univ. of Arizona Press), 451
\bibitem[Garufi et al.(2019)]{Garufi2019} Garufi, A., Podio, L., Bacciotti, F., et al.\ 2019, \aap, 628, A68
\bibitem[Grankin et al.(2007)]{Grankin2007} Grankin, K.~N., Melnikov, S.~Y., Bouvier, J., Herbst, W., \& Shevchenko, V.~S.\ 2007, \aap, 461, 183 
\bibitem[Greene \& Lada(1996)]{Greene1996} Greene, T.~P. \& Lada, C.~J.\ 1996, \aj, 112, 2184
\bibitem[Hamann(1994)]{Hamann1994} Hamann, F.\ 1994, \apjs, 93, 485
\bibitem[Hamann \& Persson(1992)]{HP92} Hamann, F., \& Persson, S.~E.\ 1992, \apj, 394, 628 
\bibitem[Hartigan, Edwards, \& Ghandour(1995)]{HEG95} Hartigan, P., Edwards, S., \& Ghandour, L.\ 1995, \apj, 452, 736
\bibitem[Hartmann \& Kenyon(1996)]{HK1996} Hartmann, L., \& Kenyon S. J. 1996, \araa, 34, 207
\bibitem[Herbst et al.(1994)]{Herbst1994} Herbst, W., Herbst, D.~K., Grossman, E.~J., \& Weinstein, D.\ 1994, \aj, 108, 1906 
\bibitem[Isella et al.(2010)]{Isella2010} Isella, A., Carpenter, J.~M., \& Sargent, A.~I.\ 2010, \apj, 714, 1746 
\bibitem[Kafka(2015)]{Kafka2015} Kafka, S., 2015, Observations from the AAVSO International Database, http://www.aavso.org
\bibitem[Kepner et al. (1993)]{Kepner1993} Kepner, J., Hartigan, P., Yang, C., \& Strom, S. 1993, \apj, 415, L119
\bibitem[Kimura, Kawabe, \& Saito(1996)]{Kimura1996} Kitamura, Y., Kawabe, R., \& Saito, M. 1996, \apj, 457, 277
\bibitem[K{\"o}nigl \& Salmeron(2011)]{KS2011} K{\"o}nigl A., Salmeron R., 2011, in Garcia P., ed., Physical Processes in Circumstellar Disks Around Young Stars. University of Chicago Press, Chicago, Chap. 7, pp 283-354
\bibitem[Lavalley-Fouquet, Cabrit, \& Dougados(2000)]{Lavalley2000} Lavalley-Fouquet, C., Cabrit, S., \& Dougados, C.\ 2000, \aap, 356, L41
\bibitem[Lavalley et al.(1997)]{Lavalley1997} Lavalley, C., Cabrit, S., Dougados, C., Ferruit, P., \& Bacon, R.\ 1997, \aap, 327, 671 
\bibitem[McGregor et al.(2002)]{McGregor2002} McGregor, P., et al. 2002, Proc. SPIE, 4841, 178
\bibitem[Maurri et al.(2014)]{Maurri2014} Maurri, L., Bacciotti, F., Podio, L., et al.\ 2014, \aap, 565, AA110
\bibitem[Mundt et al.(1987)]{Mundt1987} Mundt, R., Brugel, E. W., \& B{\"u}hrke, T. 1987, \apj, 319, 275
\bibitem[Mundt \& Fried (1983)]{Mundt1983} Mundt, R., \& Fried, J. W. 1983, \apj, 274, L83
\bibitem[Muzerolle, Hartmann, \& Carvet(1998)]{Muzerolle1998} Muzerolle, J., Hartmann, L., \& Carvet, N.\ 1998, \aj, 492, 743 
\bibitem[Oh et al.(2015)]{Oh2015}  Oh, H., Pyo, T. -S., Yuk, I. -S., Park, B. -G.\ 2015, JKAS, 48, 113 
\bibitem[Petrov \& Kozack(2007)]{Petrov2007} Petrov, P.~P., \& Kozack, B.~S.\ 2007, Astronomy Reports, 51, 500 
\bibitem[Pyo et al.(2003)]{Pyo03} Pyo, T. -S., Kobayashi, N., Hayashi, M., et al.,\ 2003, \apj, 590, 340
\bibitem[Pyo et al. (2009)]{Pyo09} Pyo, T. -S., Hayashi, M., Kobayashi, N., Terada, H., \& Tokunaga, A. T.\ 2009, \apj, 694, 654
\bibitem[Pyo et al.(2014)]{Pyo14} Pyo, T. -S., Hayashi, M., Tracy, L. B., Christopher, J. D., \& Takami, M.\ 2014,  \apj, 786, 63 
\bibitem[Pudritz et al.(2007)]{Pudritz2007} Pudritz, R.~E., Ouyed, R., Fendt, C., Brandenburg, A.\ 2007, in Reipurth, B., Jewitt, D., Keil, K. Eds. Protostars and planets V. University of Arizona Press, Tucson, pp 277-294
\bibitem[Rodr\'{i}guez et al.(2012)]{Rod2012} Rodr\'{i}guez, L. F., Gonz\'{a}lez, R. F., Raga, A. C., et al. 2012, \aap, 537, A123
\bibitem[Romanova et al.(2009)]{Romanova2009} Romanova, M.~M., Ustyugova, G.~V., Koldoba, A.~V., \& Lovelace, R.~V.~E.\ 2009, \mnras, 399, 1802 
\bibitem[Shang, Shu, \& Glassgold(1998)] {Shang1998} Shang, H., Shu, F.~H., \& Glassgold, A.~E.\ 1998, \apj, 493, L91
\bibitem[Shu et al.(1994)]{Shu1994} Shu, F., Najita, J., Ostriker, E., et al.\ 1994, \apj, 429, 781
\bibitem[Shu et al.(1995)]{Shu1995} Shu, F., Najita, J., Ostriker, E., \& Shang, H.\ 1995, \apj, 455, L155
\bibitem[Shu et al.(2000)]{Shu2000} Shu, F.~H., Najita, J.~R., Shang, H., Li, Z. -Y. 2000, in Mannings, V., Boss, A., Russell, S. Eds., Protostars and Planets IV, University of Arizona Press, Tucson, pp 789--814
\bibitem[Solf \& Bohm(1993)]{SB93} Solf, J., \& B{\"o}hm, K. H. 1993, \apj, 410, L31
\bibitem[Takami et al.(2002)]{Takami2002} Takami, M., Chrysostomou, A., Bailey, J., et al.\ 2002, \apjl, 568, L53 
\bibitem[Takami et al.(2020)]{Takami2020} Takami, M., Beck, T.~L., Schneider, P.~C., et al.\ 2020, \apj, 901, 24.
\bibitem[Takami et al.(2023)]{Takami2023} Takami, M., G\"nther, H. M., Schneider, P. C., et al.\ 2023, \apjs, 264, 1
\bibitem[Terquem et al.(1999)]{Terq1999} Terquem, C., Eisl\"offel, J., Papaloizou, J.~C.~B., \& Nelson, R. P.\ 1999, \apj, 512, L131 
\bibitem[Vrba et al.(1993)]{Vrba1993} Vrba, F.~J., Chugainov, P.~F., Weaver, W.~B., \& Stauffer, J.~S.\ 1993, \aj, 106, 1608 
\bibitem[Whelan, Ray, \& Davis(2004)]{Whelan2004} Whelan, E.~T., Ray, T.~P., \& Davis, C.~J.\ 2004, \aap, 417, 247
\bibitem[White et al.(2014)]{White2014_1} White, M.~C., McGregor, P.~J., Bicknell, G.~V., Salmeron, R., \& Beck, T.~L.\ 2014, \mnras, 441, 1681 
\bibitem[Zucker et al.(2019)]{Zucker2019} Zucker, C., Speagle, J., Schelafly, E., et al.\ 2019, \apj, 879, 125
\end{thebibliography}
\end{document}